\documentclass[11pt]{article}
\usepackage{jcapmod}

\usepackage{booktabs}
\usepackage[english]{babel}
\usepackage{amsmath,amssymb,amsbsy,amstext, amsthm, simplewick}
\usepackage{hyperref}
\usepackage{graphicx}
\usepackage{amsfonts}
\usepackage{amssymb}
\usepackage[small]{caption}
\usepackage{upgreek}
 \usepackage{exscale,relsize}

\usepackage{colortbl}
\definecolor{lightgreen}{cmyk}{0.2, 0, 0.2, 0.2}
\definecolor{lightgray}{cmyk}{0.1,0.2,0,0.1}
\definecolor{lightgray2}{cmyk}{0.1,0.1,0,0.1}

\makeatletter
\newlength{\apb@width}
\newcommand{\autoparbox}[2][c]{\settowidth{\apb@width}{#2}\parbox[#1]{\apb@width}{#2}}

\makeatother

\numberwithin{equation}{section}

\def\beq{\begin{equation}}
\def\eeq{\end{equation}}

\def\bea{\begin{eqnarray}}
\def\eea{\end{eqnarray}}

\def\D{\nabla}
\def\d{{\rm d}}

\def\beq{\begin{equation}}
\def\eeq{\end{equation}}
\def\bea{\begin{eqnarray}}
\def\eea{\end{eqnarray}}

\def\D{{\cal D}}
\def\Db{\bar{{\cal D}}}
\def\d{{\rm d}}

\def\Df{\bar{ D}}

\def\Phid{{\bar \Phi}}
\def\phid{{\bar \phi}}
\def\Rd{{\bar R}}
\def\Md{{\bar M}}
\def\Fd{{\bar F}}
\def\dalpha{\dot{\alpha}}
\def\dbeta{\dot{\beta}}
\def\ddelta{\dot{\delta}}
\def\dgamma{\dot{\gamma}}

\def\Mp{M_{\rm pl}}
\def\d{{\rm d}}

\DeclareRobustCommand{\SkipTocEntry}[4]{}

\begin{document}

\begin{titlepage}

\setcounter{page}{1} \baselineskip=15.5pt \thispagestyle{empty}

\bigskip\

\vspace{2cm}
\begin{center}
{\fontsize{20}{32}\selectfont  \bf Supergravity for Effective Theories}
\end{center}

\vspace{0.5cm}
\begin{center}
{\fontsize{14}{30}\selectfont   Daniel Baumann$^{\clubsuit,\heartsuit}$ and Daniel Green$^\clubsuit$}
\end{center}


\begin{center}
\vskip 8pt
\textsl{$^\clubsuit$ School of Natural Sciences,
 Institute for Advanced Study,
Princeton, NJ 08540, USA}

\vskip 7pt
\textsl{$^\heartsuit$ D.A.M.T.P., Cambridge University, Cambridge, CB3 0WA, UK}

\end{center}

\vspace{1.2cm}
\hrule \vspace{0.3cm}
{ \noindent \textbf{Abstract} \\[0.2cm]
\noindent 
Higher-derivative operators are central elements of any effective field theory. In supersymmetric theories, these operators include terms with derivatives in the K\"ahler potential. 
We develop a toolkit for coupling such supersymmetric effective field theories to supergravity.  We explain how to write the action for minimal supergravity coupled to chiral superfields with arbitrary numbers of derivatives and curvature couplings.  We discuss two examples in detail, showing how the component actions agree with the expectations from the linearized description in terms of a Ferrara-Zumino multiplet.
In a companion paper~\cite{Paper2}, we apply the formalism to the effective theory of inflation.}
 \vspace{0.3cm}
 \hrule       

\vspace{0.6cm}
\end{titlepage}

\tableofcontents

\newpage
\section{Introduction}

Supersymmetry (SUSY) \cite{SUSY} is an attractive framework for explaining the radiative stability of scalar masses.
Since scalar fields play fundamental roles in high-energy physics and cosmology, supersymmetry is often of critical importance in constructing theories that respect basic notions of technical naturalness \cite{'tHooft:1979bh}.
Moreover, gravity exists, so if supersymmetry is realized in Nature, we are led to consider theories of supergravity (SUGRA).

\vskip 4pt
Few concepts in theoretical physics are more widely applicable than effective field theory~(EFT)~\cite{Weinberg:1995mt}.
Progress in diverse fields, such as condensed matter physics, particle physics, and cosmology, has relied heavily on the power and universality of EFT techniques.
EFT isolates the relevant low-energy 
degrees of freedom, while systematically including the effects of high-energy  
degrees of freedom as non-renormalizable corrections.
The low-energy physics is then described by an effective action for the light fields that includes all operators that are consistent with the symmetries of the problem.  This list will include operators with arbitrary numbers of derivatives.  

\vskip 4pt
In this paper, we develop the fundamental tools for constructing effective field theories in supergravity.
Specifically, we will show how to couple higher-derivative terms in the K\"ahler potential to supergravity.
Although the literature on higher-derivative matter coupled to minimal supergravity is limited~\cite{Cecotti:1986jy,Cecotti:1986pk,Krasnikov:1989ry,Poppitz:1990uj, Butter:2010jm}, the techniques we will use are standard and well-known to supergravity experts.
Our goal here is to provide an algorithmic approach that other effective field theorists may follow to consistently couple any EFT of interest to supergravity.
In a companion paper~\cite{Paper2}, we will apply this formalism to the effective theory of inflation~\cite{EFT1, EFT2}.

\vskip 4pt
Standard treatments of minimal supergravity (e.g.~\cite{WB}) are usually formulated for theories with arbitrary K\"ahler potential $K(\Phi, \Phid)$.  Although this includes many important operators in an EFT, it is restricted to the lowest order in derivatives in the action.
However, in many effective theories the first important corrections arise from operators with higher derivatives.  This is especially true for theories of spontaneous symmetry breaking.  In that case, the two-derivative action for the Goldstone degree of freedom is typically universal, and any non-universal behavior is encoded in higher-derivative terms.  For example, the effective action for the Goldstino has an exact $R$-symmetry unless higher-derivative terms are included in the K\"ahler potential \cite{Komargodski:2009rz}. 
This further motivates understanding the coupling of higher-derivative operators to supergravity.

\vskip 4pt
The outline of the paper is a follows: In Section~\ref{sec:method}, we describe our approach for coupling higher-derivative theories to supergravity. This formalism is then applied to various examples:
In Section~\ref{sec:example}, we derive the component action for the simplest higher-derivative K\"ahler potential. Besides illustrating our approach, the results of this example are fundamental building blocks for the construction of more complex effective theories. We also check that our results are consistent with a linearized coupling of gravity to a Ferrera-Zumino multiplet.  In Section~\ref{sec:application}, we present an application to the supersymmetric effective theory of inflation~\cite{EFT1,EFT2}.
We derive the basic component actions, but defer a detailed physics discussion to a companion paper~\cite{Paper2}. We state brief conclusions in Section~\ref{sec:conclusion}.
Two technical appendices form an integral part of this work: In Appendix \ref{sec:toolkit}, we cite key identities that proved to be invaluable for the computations presented in this paper. In Appendix \ref{sec:proofs}, we give detailed derivations of some of the less trivial identities.
Unless indicated otherwise, our notation and conventions will follow Wess and Bagger~\cite{WB}.

\section{Minimal Supergravity with Higher Derivatives}
\label{sec:method}
Our starting point are four-dimensional effective actions with global ${\cal N} = 1$ supersymmetry. We allow for an arbitrary number of chiral superfields\footnote{For clarity, we will write all actions for a single chiral field $\Phi$, but the extension to multiple fields $\Phi_i$ is straightforward.  Moreover, there is no obstacle to including vector superfields when the rigid theory has a well-defined FZ multiplet.}   
 and derivatives\footnote{For reasons that we will explain in \S\ref{sec:sup}, we will not include derivatives in the superpotential.}
\beq\label{eqn:rigid}
S_{\rm rigid} = \int \d^4 x \left[\, \int d^4 \theta\, K(\Phi, \Phid; D_{\alpha} \Phi,\Df_{\alpha} \Phid; \partial_\mu \Phi,\partial_\mu \Phid; \cdots) + \left( \int d^2 \theta\, W(\Phi) + {\rm h.c.} \right) \right] \ .
\eeq
In this section, we give a brief review of our approach for coupling this higher-derivative theory to supergravity.
We will follow closely the standard methods described in Wess and Bagger's book~\cite{WB}.
Their specific treatment has the advantage that it leads to an algorithmic method for coupling any theory of chiral superfields, such as eq.~(\ref{eqn:rigid}), to supergravity.\footnote{The compensator approach to supergravity (e.g. \cite{Gates:1983nr}) may have all the same features in the hands of experts, but we found the approach of \cite{WB} to be more intuitive to the novice.  Nevertheless, these alternative techniques may be advantageous when the couplings of interest arise predominantly through the compensator \cite{Randall:1998uk, Cheung:2011jp}.} 

\subsection{Supergravity in Curved Superspace}

Our goal is to define effective actions that are both covariant and supersymmetric.
We will do this in superspace.
Since supersymmetry is a spacetime symmetry, the transformation properties of fields will depend on the metric.  Following \cite{WB}, we modify the definition of superspace and the SUSY transformations to be invariant in the presence of a dynamical metric (and the gravitino).  In this subsection, we will briefly review how covariant superfields and derivatives are defined. In the next subsection, we describe how to use curved superspace to construct invariant supergravity actions. 

\vskip 4pt
Much of the structure of curved superspace is required because a supersymmetric theory necessarily contains spinors (fermions) living in a dynamical spacetime.  In order to describe spinors in general relativity, one introduces a vierbein (or tetrad) $e_{\mu}^a(x)$, with $g_{\mu \nu} = e_\mu^a e_\nu^b \eta_{ab}$. Here, $\mu$ is a spacetime index\footnote{Here, we differ from the notation of Wess and Bagger, as we will label spacetime vector indices as $\mu$ and spacetime spinor indices as $\underline{\alpha}$.} and $a$ is a Lorentz structure index.  The tetrad formalism allows us to translate spinors that are naturally defined in flat spacetime to curved spacetime through the projection operator $\bar \sigma^{\alpha \dalpha}_\mu = e^a_\mu \bar \sigma^{\alpha \dalpha}_a$, where the matrices $\bar\sigma_a$ satisfy the Clifford algebra.  
Covariant derivatives $\D_\mu$ are then defined in terms of the spin connection $\omega_{\mu}^{ a b}$, such that $\D_\mu V^a = \partial_\mu V^a + \omega^{a}_{\mu b} V^b$. A priori, the spin connection and the vierbein are not related---however, they may be related by demanding that the connection is torsion free, i.e.~$T_{\mu \nu}^a = \D_\mu e_\nu^a - \D_\nu e_\mu^a = 0$.  Finally, for vanishing torsion we may define the curvature tensor via $[ \D_\mu, \D_\nu ] V^a = R_{\mu \nu}{}^a {}_b V^b$.

 Superspace is an extension of spacetime that includes fermionic coordinates, 
 \beq
 z^{M} \equiv \{ x^{\mu}, \theta^{\underline{\alpha}}, \bar \theta_{ \underline{\dalpha}} \} \ .
 \eeq  
 We therefore must extend our definitions of the vierbein and the spin connection to be covariant with respect to superspace.  We are led to define a new vielbein $E_M^A(z)$ and a new connection $\Omega_{M}^{ A B}(z)$, where the indices $M, A, B$ run over both vector and spinor indices.
 This implies the following definition for covariant derivatives with respect to the structure index
 \beq
\D_A V_B= \partial_A V_B - E^M_A \Omega_{M B}^C V_C \ .
\eeq
As usual, one can project this onto derivatives with respect to the spacetime index by using the vielbein $\D_M \equiv E^A_M \D_A$.  Given the covariant derivatives, we can define the torsion 
\beq
T_{NM}^A = \D_N E_M^A - (-1)^{nm} \D_M E_N^A \ , 
\eeq
where $n$, $m$ are $1$ ($0$) when the index is a spinor (vector).  The following identity holds: $T_{NM}^A = (-1)^{n(m+b)} E_M^B  E_{N}^C T_{CB}^A $. The superspace curvature tensor is
\beq
R_{NM}{}_C{}^D = \D_N \Omega_{MC}^D - (-1)^{nm} \D_M \Omega_{N C}^D  +(-1)^{n(m+c+e)} \Omega_{MC}^E \Omega_{NE}^D-(-1)^{m(c+e)} \Omega_{NC}^E \Omega_{ME}^D\ .
\eeq
These three operators are related through an extremely useful identity
\beq
\label{equ:commutator}
\left(\D_{C} \D_B - (-1)^{bc} \D_B \D_C \right) V^{A} = (-1)^{d(c+b)} V^D  R_{CB D}{}^A - T_{CB}^D \D_D V^A \ .
\eeq
We will use eq.~(\ref{equ:commutator}) over and over.
Like in the non-supersymmetric case, we wish to determine $\Omega_{M}^{ A B}$ in terms of derivatives of $E_M^A$.  However, in the context of SUSY, the torsion cannot vanish.  This is clear from eq.~(\ref{equ:commutator}) if we set $C = \alpha$ and $B = \dalpha$. In this case, the components of the torsion must be chosen to reproduce the SUSY algebra, $T_{\alpha \dalpha}^a = 2 i\sigma_{\alpha \dalpha}^{a}$.  In general, one therefore has to solve the Bianchi identities to relate the components of $E_M^A$ and $\Omega_{M}^{ A B}$.  In this paper, we will study minimal supergravity, which represents one self-consistent set of solutions to these identities---but, as the name suggests, it is not the most general solution.

The full solution to the Bianchi identities, given the minimal SUGRA ansatz, can be found in~\cite{WB, Grimm:1978ch,Dragon:1978nf,Girardi:1984eq}. For the convenience of the reader we reproduce some of the more important results in Appendix~\ref{sec:toolkit}.  The non-zero torsion components will be particularly important and are given by
\begin{align}
T_{\alpha \dot \alpha}^{a} &\ =\ 2i \sigma_{\alpha \dot \alpha}^a \ , \\
T_{\dot \beta a}^{\alpha}&\ =\ - i \sigma_{\dot \beta a}^{\alpha} R\ , \label{equ:TR} \\
T_{\beta a}^{\alpha} &\ =\ \tfrac{i}{8} \bar\sigma_a^{\gamma \dot\gamma}(\delta_{\gamma}^{\alpha} G_{\beta \dot\gamma} - 3 \delta_{\beta}^{\alpha} G_{\gamma \dot \gamma} + 3 \epsilon_{\beta \gamma} G^{\alpha}_{\dot\gamma})\ , \label{equ:TG} 
\end{align}
where $R$---the superspace curvature---is a chiral superfield and $G_{\alpha \dot \alpha}^\dag = G_{\alpha \dot \alpha}$. Here, we used the standard notation $\bar \sigma^{\alpha \dot \alpha}_a \equiv \epsilon^{\dot \alpha \dot \beta} \epsilon^{\alpha \beta} \sigma_{a \beta \dot \beta}$.
Expressions with dotted and undotted indices interchanged are obtained by complex conjugation.  All other components of the torsion vanish.

The final step in producing a dynamical gravity multiplet is to eliminate unphysical modes by gauge fixing.  The gauge symmetries of supergravity that we wish to maintain are general coordinate transformations, local SUSY transformations and local Lorentz (structure) transformations.  We have promoted these to local superspace transformations.  Therefore, we will leave the $\theta =\bar \theta =0$ components unfixed, but gauge fix the higher components of these transformations.  We choose the lowest components of the vielbein to take the form
\beq
E_M^A\big|_{\theta = \bar \theta = 0} =  \left( \begin{array}{ccc}
e_\mu^a & \tfrac{1}{2} \psi_\mu^\alpha & \tfrac{1}{2} \bar \psi_{\mu \dalpha} \\
0 & \delta_{\underline{\alpha}}^{\alpha} & 0 \\
0 & 0 & \delta^{\underline{\dalpha}}_{\dalpha} \end{array} \right)\ , \label{equ:VIEL}
\eeq
where $ \psi_\mu^\alpha$ is the gravitino\footnote{Thinking of supergravity as a gauge theory, we would arrive at the gravitino as the gauge field corresponding to the SUSY generators $Q^\alpha$.} 
and $\underline{\alpha}, \underline{\dot \alpha}$ are spacetime spinor indices.  Even after gauge fixing, the lowest components of $R$ and $G_a \equiv - \tfrac{1}{2} \bar \sigma^{\alpha \dalpha}_a G_{\alpha \dalpha} $ are undetermined.  Therefore, we introduce auxiliary fields $M$ and $b_a$~\cite{Stelle:1978ye,Ferrara:1978em}, such that 
\beq
R|_{\theta =\bar\theta =0} = -\tfrac{1}{6} M \qquad {\rm and} \qquad G_a|_{\theta =\bar \theta = 0} = -  \tfrac{1}{3} b_{a}\ .
\eeq  
All other components can then be determined by the solutions to the Bianchi identities.  

\vspace{4pt}
In summary, minimal supergravity contains two dynamical fields, the metric $e_\mu^a$ and the gravitino $\psi_\mu^\alpha$, and two auxiliary fields, a complex scalar $M$ and a real vector $b_a$.  All components of the vielbein, spin connection, torsion and curvature tensors can be determined in terms of these fields and their derivatives.  A list of useful identities can be found in Appendix~\ref{sec:toolkit} and some instructive derivations are presented in Appendix~\ref{sec:proofs}.

\subsection{Actions for Supergravity in Superspace}
\label{sec:sup}

Having introduced the superspace formulation of the supergravity multiplet, we will now show how to construct  actions that couple supergravity to chiral superfields.

\vskip 4pt
In flat space, we define a superfield $\Phi$ as chiral if $\Df_{\dalpha} \Phi = 0$.  The obvious generalization to curved space is to define a chiral superfield by $\Db_{\dalpha} \Phi = 0$.  The components of $\Phi$ are then defined as
\beq
\phi =  \Phi \big|_{\theta =\bar \theta = 0} \ , \qquad
\chi_{\alpha} =  \tfrac{1}{\sqrt{2}}\D_{\alpha} \Phi \big|_{\theta =\bar \theta = 0} \ ,  \qquad {\rm and} \qquad
F =  -\tfrac{1}{4}  \D^2 \Phi \big|_{\theta =\bar \theta = 0} \ . \label{equ:comp}
\eeq
It is important to note that $\Db_{\dalpha} \D_a \Phi \neq 0$.  Therefore, in supergravity, {\it the derivative of a chiral field is not a chiral field}. This simple fact is responsible for much of the complexity of higher-derivative supergravity.

It is convenient to define a new set of superspace coordinates, $\Theta_{\alpha}$, in terms of which the component expansion of chiral fields is 
\beq 
\Phi = \phi + \sqrt{2} \Theta \chi + \Theta^2 F \ .
\eeq
These coordinates allow us to construct chiral superspace integrals as $\int d^2 \Theta \, E \, W(\Phi)$, where $W(\Phi)$ is an arbitrary holomorphic function of $\Phi$.  For this to be invariant, we need to choose the chiral density $E$ such that the SUSY variation of the integral is a total derivative.  
In these coordinates, one finds that 
\beq
E = e \left\{1 + i\,  \Theta \sigma^a \bar \psi_{a} - \Theta^2 \left[  \bar M + \tfrac{1}{4}  \bar \psi_a (\bar\sigma^{a} \sigma^{b} - \bar \sigma^b \sigma^a) \bar \psi_b \right]  \right\}\ ,
\eeq
where $e \equiv \det e^a_\mu = \sqrt{-g}$.
Supersymmetric contributions to the action can therefore be written as $\int \d^4 x  \int d^2 \Theta \, E \, W( \Phi)$.

In order to construct kinetic terms or higher-derivative operators, we must be able to write an action that includes non-chiral fields.  Having defined the appropriate chiral measure, we just need to project operators onto chiral operators.  In flat space, given any superfield $\mathcal{O}$, we can construct a chiral operator via $\Db^2 \mathcal{O}$.  In curved space, the situation is modified because $\Db_{\dot \alpha} \Db^2 {\cal O} \neq 0$.  However, using eq.~(\ref{equ:commutator}), one can show that 
\beq
\Db_{\dalpha} (\Db^2 - 8 R) \mathcal{O} = 0 \ ,
\eeq
where the operator $\mathcal{O}$ is a Lorentz scalar, but otherwise arbitrary.  Hence, the operator $(\Db^2 - 8 R) \mathcal{O}$ is chiral and we can use the chiral superspace to define invariant actions.
The two-derivative action for minimal supergravity can then be written as~\cite{WB}
\beq
\label{equ:S0}
S_0 = \frac{1}{\kappa^2} \int \d^4 x \int d^2 \Theta \, E \left[ \tfrac{3}{8} (\Db^2 - 8 R) \, e^{-\frac{\kappa^2}{3}K(\Phi, \Phid)} + \kappa^2 W(\Phi)\right] + {\rm h.c.}
\eeq
where $\kappa \equiv 1/\Mp$.  
However, note that the statement that $(\Db^2 - 8 R) \mathcal{O}$ projects $\mathcal{O}$ onto a chiral field only requires $\mathcal{O}$ to be a Lorentz scalar.  We may therefore include covariant derivatives $\D_a$, $\D_\alpha$ and $ \Db_{\dalpha}$ in ${\cal O}$, as long as all the indices are contracted with the appropriate metric, i.e.~$\eta_{ab}, \epsilon_{\alpha \beta}$ or $\epsilon_{\dalpha \dbeta}$.\footnote{We should also point out that there is nothing that prevents us from including explicit couplings to curvature inside the K\"ahler potential.  We may also include functions of $G_{a}$, $R$, $R_{a b}$, etc.~as long as the indices are contracted.  The chiral projector ensures that all such terms are supersymmetric.  Although the couplings to $R$ could easily have been included in two-derivative theories, they can safely be ignored as they only contribute at fourth order (and higher) in the derivative expansion.  In higher-derivative theories, the role of curvature couplings will be more prominent. We will discuss explicit curvature couplings in a concrete example in Section~\ref{sec:example}.}
There are also circumstances where additional derivatives in the superpotential can be important (e.g. \cite{Beasley:2005iu, Antoniadis:2007xc}).  However, since $\D_A \Phi$ is no longer chiral in supergravity, these couplings require additional terms to cancel non-zero supersymmetry transformations.  We will choose to avoid this complication by assuming that higher-derivative terms originate purely from the K\"ahler potential and that the superpotential is simply $W(\Phi)$.  
\vspace{8pt}

In summary, we may write the superspace action for minimal supergravity coupled to any theory of chiral superfields as
\beq
\label{equ:higherS}
S =  \frac{1}{\kappa^2} \int \d^4 x \int d^2 \Theta \, E \left[ \tfrac{3}{8} (\Db^2 - 8 R) \, e^{ -\tfrac{\kappa^2}{3}K(\Phi, \Phid; \D_{\alpha} \Phi,\Db_{\alpha} \Phid; \D_a \Phi,\D_a \Phid; R, \bar R, G_a; \cdots)}+ \kappa^2 W(\Phi)\right] + {\rm h.c.}
\eeq
In the rigid limit, $\kappa \to 0$, this action reproduces eq.~(\ref{eqn:rigid}).

\subsection{Actions for Supergravity in Components}

So far, we have reviewed how to construct manifestly supersymmetric and covariant actions in superspace.  To discuss the consequences of the theory, we will need to determine the action in components.

\vskip 4pt
At lowest order in $\kappa$, eq.~(\ref{equ:S0}) includes the invariant action for the supergravity multiplet
\beq
S_{\rm s.g.} = - \frac{3}{\kappa^2}  \int \d^4 x \int d^2 \Theta \, E R \ =\,  \frac{1}{\kappa^2} \int \d^4 x \, \sqrt{-g} \left[ - \tfrac{1}{2} {\cal R} - \tfrac{1}{3} |M|^2 + \tfrac{1}{3} b^2 \right]  \ +\ {\rm fermions} \ ,
\eeq
where we used the component expansion of the superspace curvature,
\beq
R = - \tfrac{1}{6} \left\{ M + \Theta^2 \left(-\tfrac{1}{2}{\cal R} + \tfrac{2}{3}|M|^2 + \tfrac{1}{3} b_a b^a - i\D_a b^a \right) \right\} \ +\ {\rm fermions}\ .
\eeq
At higher order in $\kappa$, eq.~(\ref{equ:S0}) leads to the standard kinetic term and the F-term potential for the scalar $\phi$~\cite{WB}.
Writing the higher-derivative action (\ref{equ:higherS}) in component form is conceptually straightforward, but often requires tedious calculations.  The basic challenge is to determine the $\Theta$-expansion of a given operator after it has been projected onto chiral superspace.  Given a general chiral operator $(\Db^2-8R)\mathcal{O}$, its components are found by acting with spinor derivative and taking lowest components,
\beq\label{eqn:expansion}
(\Db^2-8R)\mathcal{O} =\left. (\Db^2-8R)\mathcal{O} \right|\, +\, \left. \Theta^{\alpha} \D_{\alpha} \left[ (\Db^2-8R)\mathcal{O} \right] \right| \left. \, -\, \tfrac{1}{4} \Theta^2 \D^2 \left[  (\Db^2-8R)\mathcal{O} \right] \right| \ .
\eeq
Here and in the following, $(\ldots)|$ is shorthand for $(\ldots)|_{\theta=\bar \theta =0}$.
In the case of interest, $\mathcal{O}$ is a function of chiral and anti-chiral superfields with or without derivatives acting on them.  Through repeated application of eq.~(\ref{equ:commutator}) we can always reduce any given term to components of the chiral field plus couplings to curvature and torsion elements. Determining these couplings
is one of our primary tasks.

\vskip 4pt
In order to gain some intuition, consider the problem of finding the components of the operator $(\bar \D^2 - 8 R) \mathcal{O}$, where $\mathcal{O} = \D_a \Phi\, \mathcal{O}^a$. This requires knowledge of all possible fermionic derivatives acting on $\D_a \Phi$, cf.~eq.~(\ref{eqn:expansion}).  Let us illustrate this with two simple examples:
\begin{itemize}
\item
First, we consider the term $\D_{\alpha} \D_a \Phi |$.  In flat space, this problem is trivial because $\D_{\alpha}$ commutes with $\D_a$.  However, in curved space $\D_{\alpha}$ and $\D_a$ don't commute and the problem is slightly more involved.
Applying eq.~(\ref{equ:commutator}) and using eqs.~(\ref{equ:TG}) and (\ref{equ:VIEL}), we find \begin{align}
\D_{\alpha} \D_a \Phi \big| &= \D_a \D_{\alpha} \Phi | + [\D_\alpha, \D_a ] \Phi| \nonumber \\
&=  \D_a  \D_{\alpha} \Phi| - T_{\alpha a}^{\beta}| \D_{\beta} \Phi|  \nonumber \\
&= E^M_a \D_M \D_{\alpha} \Phi| -  T_{\alpha a}^{\beta}| \D_{\beta} \Phi|  \nonumber \\
&= e^{\mu}_a\left[ \sqrt{2}   \nabla_{\mu} \chi_{\alpha} + \tfrac{1}{2} \psi_\mu^{\beta} \D_{\beta} \D_{\alpha} \Phi| \right] - T_{\alpha a}^{\beta}| \D_{\beta} \Phi|  \nonumber \\
&= e^{\mu}_a \left[ \sqrt{2}  \nabla_{\mu} \chi_{\alpha} +  \psi_{\mu \alpha} F +  i \tfrac{\sqrt{2}}{24} \bar\sigma_\mu^{\gamma \dot\gamma}(\delta_{\gamma}^{\beta} b_{\alpha \dot\gamma} - 3 \delta_{\alpha}^{\beta} b_{\gamma \dot \gamma} + 3 \epsilon_{\alpha \gamma} b^{\beta}_{\dot\gamma})  \chi_{\beta} \right] \ .
\end{align}
If we set the gravitino $\psi_{a \alpha}$ and the auxiliary field $b_a$ to zero, we reproduce the simple flat space result.

\item For a slightly more complicated example, let us determine some of the higher derivatives of $\D_a \Phi$.
By definition $\Db_{\dalpha} \Phi = 0$, so acting with $\Db_{\dalpha}$ on $\D_a \Phi$, gives  
\beq
\label{equ:1X}
\Db_{\dalpha} \D_a \Phi = [\Db_{\dalpha}, \D_a] \Phi = -  i T_{\dalpha a}^\beta \D_\beta \Phi= i R \sigma_{\dalpha a}^\beta \D_\beta \Phi\ ,
\eeq
where we used (\ref{equ:commutator}) and (\ref{equ:TR}).
Acting with another $\Db_{\dalpha}$, results in 
\beq
\label{equ:2X}
-\tfrac{1}{4} \Db^2 \D_a \Phi = R \D_a \Phi\ ,
\eeq
where we used the fact that $\Db_{\dalpha} R = 0$.
Note that the expressions (\ref{equ:1X}) and (\ref{equ:2X}) are true without restricting to the lowest components.
To determine, for instance, the term $\tfrac{1}{16} \D^2 \Db^2 \D_a \Phi |$, we take two additional derivatives
\bea
\tfrac{1}{16} \D^2 \Db^2 \D_a \Phi | &=& -\tfrac{1}{4} \left[ \D^2 R| \D_a \Phi| + 2 \D^{\alpha} R| \D_\alpha \D_a \Phi| + R| \D^2 \D_a \Phi| \right]  \\
&=&  e^{\mu}_a \left[ (\tfrac{1}{6}{\cal R} - \tfrac{5}{18}|M|^2 - \tfrac{1}{9} b^2 +  \tfrac{i}{3}\nabla \hskip -3pt \cdot \hskip -2pt b) \nabla_\mu \phi - \tfrac{1}{6} M \nabla_\mu F - \tfrac{i}{9} b_\mu MF\right] + {\rm fermions} \nonumber \ ,
\eea
where $\nabla \hskip -3pt \cdot \hskip -2pt b \equiv \nabla_\nu b^\nu$.
\end{itemize}

\vskip 6pt
This procedure can be repeated for arbitrarily complicated examples (see Appendix~\ref{sec:proofs}).  
Although the computations quickly become tedious, we emphasize that the formalism is completely algorithmic and can be applied to any higher-derivative term in the K\"ahler potential.  In order to do so, however, one must determine the component expansions of the curvature and torsion.  (For example, one may be required to evaluate $\D^2 \Db^2 G_{a} |$.)
All such terms can be determined from the Bianchi identities and the terms relevant for our examples are reproduced in Appendix~\ref{sec:toolkit}, along with several representative derivations in Appendix~\ref{sec:proofs}.

\section{Simplest Higher-Derivative Example}
\label{sec:example}

To illustrate the formalism of the previous section, we start with the simplest higher-derivative term in the K\"ahler potential: 
\beq
\label{equ:simpleK}
K = \frac{1}{\Lambda^2} \D_a \Phi \D^a \Phid\ .
\eeq
At the same time, the results that we will obtain for this example will serve as fundamental building blocks for constructing more complicated effective theories (see \S\ref{sec:application}).  
In the next two sections, we will for simplicity drop all fermionic components from our calculations.  While this means that we will not be able to check explicitly  that the component actions are invariant under SUSY transformations, it has the advantage that our formulas will stay manageable. We will discuss alternative consistency checks of our results in \S\ref{sec:FZ}.

\subsection{Supergravity Action in Components}

Let us set $\Lambda \equiv 1$ and absorb the integral measure $e$ into a rescaling of the chiral measure, $E \to e E$.
The leading contribution to the supergravity Lagrangian (\ref{equ:higherS}) then is
\beq
{\cal L} = \int d^2 \Theta \, E\left[ \, -  \tfrac{1}{8}(\Db^2 - 8 R) \D_a \Phi \D^a \Phid \, \right] \ +\ {\rm h.c.} \ +\ {\cal O}(\kappa^2)\ ,
\eeq
where
\begin{align}
 \int d^2 \Theta\, ER\, \D_a \Phi \D^a \Phid &\ =\  \left[ \tfrac{1}{12}{\cal R} + \tfrac{1}{18}|M|^2 - \tfrac{1}{18} b^2 \right] \D_a \Phi \D^a \Phid| \ +\ \tfrac{1}{24} M \D^2( \D_a \Phi \D^a \Phid)|  \ , \label{equ:no1} \\
-\tfrac{1}{8} \int d^2 \Theta\, E\, \Db^2( \D_a \Phi \D^a \Phid) &\ =\  \tfrac{1}{8}  \bar M \Db^2 (\D_a \Phi \D^a \Phid) |  \, +\, \tfrac{1}{32} \D^2 \Db^2(\D_a \Phi \D^a \Phid) |\ . \label{equ:no2x}
\end{align}
Note that the last term in (\ref{equ:no1}) and the first term in (\ref{equ:no2x}) are complex conjugates of each other.
To compute them,
\begin{align}
\D^2( \D_a \Phi \D^a \Phid)| &\ =\ \D^2 \D_a \Phi| \, \D^a \Phid| + \D_a \Phi|\, \D^2 \D^a \Phid| \ , 
\end{align}
we need to know $\D^2 \D_a \Phi|$ and $\D^2 \D^a \Phid|$. These terms are evaluated in Appendix~\ref{sec:toolkit}.
The last term in (\ref{equ:no2x}) is a bit more involved
\begin{align}
 \D^2 \Db^2(\D_a \Phi \D^a \Phid) | &\ =\ \D^2 \Db^2 \D_a \Phi|\, \D^a \Phid| +\D^2 \D_a \Phi|\,  \Db^2 \D^a \Phid| +\Db^2 \D_a \Phi|\, \D^2 \D^a \Phid|\nonumber \\
& \hspace{0.7cm} -\, 4\, \D^\alpha \Db_{\dalpha} \D_a \Phi|\, \D_{\alpha} \Db^{\dalpha} \D^a \Phid| + \D_a \Phi|\, \D^2 \Db^2 \D^a \Phid| \ .   \label{equ:no3}
\end{align}
This requires us to know $\D^2 \Db^2 \D_a \Phi|$ and $\D^2 \Db^2 \D^a \Phid|$, as well as $\D^\alpha \Db_{\dalpha} \D_a \Phi|$ and $\D_{\alpha} \Db^{\dalpha} \D^a \Phid|$. Again, the results may be found in Appendix~\ref{sec:toolkit}.
Combining everything, the final answer can be written as
\beq
\label{equ:finalS}
\fbox{${\cal L}_{\partial \Phi \partial \Phid} \ =\ - |\nabla^2 \phi|^2   + f |\nabla_\mu \phi|^2  +   \tfrac{1}{2} f_{\mu\nu}\,  \nabla^\mu \phi\, \nabla^\nu \phid  \  -\ \tfrac{1}{2} b_\mu j^\mu -\tfrac{1}{4} M \bar{\rm x} -\tfrac{1}{4} \bar M {\rm x} +  f_{\rm aux} $}\ ,
\eeq
where 
\begin{align}
&f \hspace{0.42cm} \ \equiv\ \tfrac{1}{3} {\cal R} \, +\, \tfrac{1}{9} |M|^2 \ , \label{equ:ff}
\\
&f_{\mu \nu} \hspace{0.12cm}\ \equiv\ - 2 {\cal R}_{\mu \nu} - \tfrac{4}{9} b_\mu b_\nu \ ,\\
&f_{\rm aux} \ \equiv\    |\nabla_\mu F|^2 + \tfrac{2}{9} |M|^2 |F|^2  + \tfrac{4}{9} b^2 |F|^2 \ . \label{equ:fauxaux}
\end{align}
Here, the terms linear in $b_\mu$ couple to 
\beq\label{equ:j}
\fbox{$\displaystyle j^\mu \ \equiv\  - \tfrac{4i}{3}\left(  \nabla^\mu \phi \nabla^2 \phid - \tfrac{1}{2} \nabla^\nu \phi \nabla_\nu \nabla^\mu \phid  + F \nabla^\mu \Fd - \tfrac{1}{3} M F \nabla^\mu \phid -  {\rm h.c.} \right)$}\ ,
\eeq
while the terms linear in $M$ couple to 
\beq
\fbox{$\displaystyle \bar {\rm x} \ \equiv\ \tfrac{4}{3} \nabla_\mu F \nabla^\mu \phid + \tfrac{4}{3} F \nabla^2 \phid $}\ . \label{equ:x}
\eeq
Writing the answer in the form of eq.~(\ref{equ:finalS}) involves integrating by parts.  For later applications, it will also be useful to include the result without any such integrations,
\begin{align}
\label{equ:nobyparts}
{\cal \tilde L}_{\partial \Phi \partial \Phid} &\ =\  {\cal L}_{\partial \Phi \partial \Phid} \nonumber \\
&\hspace{0.7cm} + \left[\, \tfrac{1}{2} \nabla_\mu \hskip -1pt \left( \nabla^2 \phi \nabla^\mu \phid  + \tfrac{1}{3} F M \nabla^\mu \phid  - \tfrac{4 i}{3} b_\nu \nabla^\nu \phi \nabla^\mu \phid\, \right) + {\rm h.c.} \, \right]\ .
\end{align}

\subsection{Comparison with Linearized Supergravity}\label{sec:FZ}

As a consistency check, we now compare our result to the generic expectations from linearized supergravity.
Let us remind the reader what these expectations are (for recent discussions see also \cite{Komargodski:2010rb, Festuccia:2011ws, Dumitrescu:2011iu}):

Given a theory for matter coupled to gravity, consider the problem of small fluctuations of the metric around flat space, $g_{\mu \nu} = \eta_{\mu \nu } + h_{\mu \nu}$, with $h_{\mu \nu} \ll 1$.  If we expand the action to linear order in $h_{\mu \nu}$, the coupling to matter must take the form $h_{\mu \nu} T^{\mu \nu}$, where $T^{\mu \nu}$ is a conserved energy-momentum tensor.
For a theory of supergravity, we may take the same flat space limit and describe the linearized coupling to gravity (see e.g.~\cite{Weinberg:2000cr}).  Of course, the metric perturbation $h_{\mu \nu}$ still couples to the energy-momentum tensor, but it now appears as the $\theta \bar \theta$ component of a real supermultiplet ${\cal H}_{\alpha \dot \alpha}$.  The stress tensor is similarly embedded in a real supermultiplet ${\cal J}^{\dalpha \alpha}$ and the linearized coupling to gravity can be written as
\beq
\mathcal{L}_{\rm linear} = \int d^4 \theta\, {\cal J}_{\alpha \dalpha} {\cal H}^{\alpha \dalpha} \ . \label{equ:linear}
\eeq
Requiring that eq.~(\ref{equ:linear}) is invariant under coordinate transformations---${\cal H}_{\alpha \dalpha} \to {\cal H}_{\alpha \dalpha} + D_{\alpha} \bar L_{\dalpha} - \Df_{\dalpha} L_{\alpha}$ (to linear order)---implies 
\beq
\int d^4\theta\, \Df^{\dalpha} {\cal J}_{\dalpha \alpha} L^{\alpha} = 0 \ .
\eeq
Next, let us assume that the stress tensor is part of a Ferrara-Zumino (FZ) multiplet, whose defining relation is 
\beq
\bar D^{\dalpha} {\cal J}_{\dalpha \alpha} = D_{\alpha} X\ , \label{equ:FZ}
\eeq 
where $X$ is a chiral superfield.  For eq.~(\ref{equ:linear}) to be invariant under coordinate transformations, now implies the more restrictive constraint $\bar D^2 D^{\alpha} L_{\alpha} = 0$.  Moreover, it can be shown that this restriction of the space of gauge transformations is equivalent to coupling to minimal supergravity.  Running the argument in reverse, we conclude that consistency of minimal supergravity {\it requires} linear couplings between the metric multiplet and an FZ multiplet. 
In components, this statement takes the form 
\beq
 {\cal L}_{\rm linear} \ \subset\ - \tfrac{1}{2} b^\mu {\cal J}_\mu| - \tfrac{1}{4} M \bar X| - \tfrac{1}{4} \Md X|\ .
\eeq
where ${\cal J}_\mu|$ and $X|$ are the lowest components of ${\cal J}_{\dalpha \alpha}$ and $X$, respectively.  
In this section, we will confirm that the linearized couplings that we found in the previous section are consistent with linear couplings to an FZ multiplet.  By satisfying (\ref{equ:FZ}), the linear couplings provide a non-trivial check not only on the form of the couplings, but also on the numerical coefficients of our computation.

\subsubsection{The FZ-Multiplet}

We would like to find the FZ-multiplet that contains the energy-momentum tensor associated with the action
\beq \label{eqn:actionrigid}
\int d^4 \theta \, \D_a \Phi \D^a\Phid \ .
\eeq
We could use the Noether procedure to determine the multiplet~\cite{Magro:2001aj, Kuzenko:2010ni}.  However, it will prove to be easier to determine the FZ multiplet directly from (\ref{equ:FZ}) by writing a linear combination of all possible terms in $J_{\dalpha \alpha}$ and $X$ and solving for their coefficients.

As usual, a good ansatz will simplify the calculation.  Since there is only one term in (\ref{eqn:actionrigid}), all terms in the stress tensor should have the same number of derivatives.  Moreover, eq.~(\ref{eqn:actionrigid}) has a symmetry under $\Phi \to \Phi + const.$, which should not be broken by the coupling to gravity.  This further constrains the ansatz for the FZ multiplet. Requiring that all terms have the same number of derivatives, ${\cal J}_{\dalpha \alpha}$ and $X$ take the following forms
\bea
{\cal J}_{\dalpha \alpha} &=& a \big( \bar D_{\dalpha} \partial_{\mu} \Phid D_\alpha \partial^{\mu} \Phi \big) + b \big( \partial_{\mu} \Phid \partial_{\dot \alpha \alpha} \partial^{\mu} \Phi \pm {\rm h.c.} \big) + c \big( \bar D_{\dalpha}  \Phid D_\alpha \partial_{\mu}\partial^{\mu} \Phi \pm {\rm h.c.} \big)  \nonumber \\
&& +\, d \big( \partial_{\dalpha \alpha}  \Phid \partial_{\mu}\partial^{\mu} \Phi \pm {\rm h.c.} \big)
+ e \big( \bar D^2 \Phid  \partial_{\dalpha \alpha}D^2 \Phi \pm {\rm h.c.} \big) \nonumber \\
&&
+\, f \big(\partial_{\beta \dbeta} \Df^{\dbeta} \Phid \partial_{\dalpha}^{\beta} D_{\alpha} \Phi+\partial_{\beta \dalpha} \Df^{\dbeta} \Phid \partial_{\dbeta}^{\beta} D_{\alpha} \Phi \pm {\rm h.c.} \big) \nonumber \\
&& +\, g \big(\partial_{\alpha \dbeta} \Df^{\dbeta} \Phid \partial_{\dalpha \beta} D^{\beta} \Phi+\partial_{\alpha \dalpha} \Df^{\dbeta} \Phid \partial_{\dbeta \beta} D^{\beta} \Phi +\partial_{\beta \dbeta} \Df^{\dbeta} \Phid \partial_{\dalpha\alpha} D^{\beta} \Phi+\partial_{\beta \dalpha} \Df^{\dbeta} \Phid \partial_{\dbeta \alpha} D^{\beta} \Phi \big)\nonumber \\
&& +\, h \big(\partial_{\alpha \dbeta}\partial_{\beta \dalpha} \Df^{\dbeta} \Phid D^\beta \Phi +\partial_{\alpha \dalpha}\partial_{\beta \dbeta} \Df^{\dbeta} \Phid D^\beta \Phi \pm {\rm h.c.} \big)\ , \label{equ:J}
\eea
and 
\beq
X = \bar D^2 \Big( p\partial_\mu \Phid \partial^{\mu} \Phi + q  (\partial_\mu \partial^\mu \Phi )\Phid \Big)\ . \label{equ:X}
\eeq
Here, the $\pm$ refer to the fact that ${\cal J}_{\dalpha \alpha}$ is real, so we have to choose the signs to be consistent.  
Substituting (\ref{equ:J}) and (\ref{equ:X}) into (\ref{equ:FZ}), we find (after some work),
\begin{align}
p = q \ , \qquad {\rm and} \qquad &a= 3h - 2	q \ , \quad
b= - 8i h + 4 i q\ , \quad
c=- 3h\ , \quad 
d= 8 i h\ , \quad \nonumber \\
& \hspace{0.5cm} 4e= 3i h - i q \ , \quad
2 f=- 3 h+q \ , \quad
2 g = - h+ 2q \ .
\end{align}
We see that this is a two parameter family of solutions---here written in terms of the parameters $q$ and $h$.
To compare this solution with (\ref{equ:j}) and (\ref{equ:x}), we compute ${\rm x}  \equiv X|$ and $j_\mu \equiv - \tfrac{1}{2} \bar \sigma^{\alpha \dalpha}_\mu {\cal J}_{\alpha \dalpha} |$, ignoring fermions, 
\begin{align}
{\rm x} &\ =\ - 4 q ( \partial_\mu \bar F \partial^\mu \phi + \bar F \partial_\mu \partial^\mu \phi )\ , \\
j_{\mu} &\ =\  -2 i \left [ -\, 2 h \partial_{\mu}  \phi \partial_{\nu}\partial^{\nu} \phid + (2h-q) \partial_{\nu} \phi \partial^{\nu} \partial_{\mu} \phid 
- (3 h - q)  F  \partial_{\mu}\bar F - {\rm h.c.} \right] \ .
\end{align}
This matches our supergravity calculation when $h = q = - \tfrac{1}{3}$. This is a non-trivial check, because we have matched the coefficients of {\it five} operators using only {\it two} variables.  We cannot match the $M$-dependent term in (\ref{equ:j}) as it arises only at non-linear order in the coupling to gravity.

\subsubsection{ Improvement Terms and Curvature Couplings}

From the definition (\ref{equ:FZ}), one might have suspected that  the form of the multiplet would be determined up to an overall normalization.  However, we found two free parameters, $h$ and $q$, that were undetermined and were only fixed by matching to the supergravity action.  This additional freedom comes from the freedom to add improvement terms to the energy-momentum tensor.  Specifically, to any energy-momentum tensor we may add a term of the form $T_{\mu \nu} \to T_{\mu \nu} + \partial^{\rho} B_{\mu \nu \rho}$, where $B_{\mu \nu \rho} = - B_{\rho \nu \mu}$.  The new energy-momentum tensor is still conserved and the conserved charges are unchanged.

When coupling a theory to gravity, the freedom to include improvement terms in the stress tensor translates into a freedom to change the curvature couplings for the theory.  For example, given a real scalar field $\phi$, we can always add the following improvement term $T_{\mu \nu} \to T_{\mu \nu} + \alpha (\partial_{\mu}\partial_{\nu} - \eta_{\mu \nu} \partial^2 )\phi^2$.  This is equivalent to the freedom to add the curvature term to the Lagrangian $\mathcal{L} \to \mathcal{L} + \alpha \phi^2 {\cal R}$.

In minimal supergravity coupled with at most two derivatives, the only freedom in the improvement terms correspond to K\"ahler transformations of the action.  However, in the higher-derivative case, we must allow a larger number of deformations.  The rigid theory described by (\ref{eqn:actionrigid}) is superconformal and must allow for an FZ multiplet with $X = 0$ \cite{Komargodski:2010rb,Dumitrescu:2011iu}.  We found this solution in the previous section, if we set $p = q = 0$.  Since we can add an improvement term to make $X=0$, we should also be able to couple to minimal supergravity without breaking conformal invariance by adding a curvature coupling.

Let us check explicitly that we can indeed add a curvature coupling in such a way that we get $X=0$ for the FZ multiplet.  Consider the following Lagrangian
\bea\label{eqn:Gcoupling}
\mathcal{L}_{\partial \Phi \partial \Phid+ G_{\alpha \dalpha}}  = \int d^2 \Theta\, E \left[ -\tfrac{1}{8} (\Db^2 - 8 R) \left( \D_a \Phi \D^a \Phid + \gamma \, G_{\alpha \dalpha} \D^{\alpha}\Phi\Db^{\dalpha}  \Phid \right) \right]\ ,
\eea
where $\gamma$ is a constant. 
We would like to find the value of $\gamma$ for which gravity couples linearly to the superconformal FZ multiplet with $X = 0$.  Because the theory is conformal, we will also check that the couplings are the conformal ones.  To make this comparison, we expand (\ref{eqn:Gcoupling}), keeping only terms linear in $M$, $b_{\mu}$ and ${\cal R}_{\mu \nu}$.  The linearized theory takes the form
\beq
\label{equ:improved}
{\cal L}^{\rm (linear)}_{\partial \Phi \partial \Phid+ G_{\alpha \dalpha}} \ =\ - |\Box \phi|^2   + f |\nabla_\mu \phi|^2  +   \tfrac{1}{2} f_{\mu\nu}\,  \nabla^\mu \phi\, \nabla^\nu \phid  \  -\ \tfrac{1}{2} b_\mu j^\mu -\tfrac{1}{4} M \bar{\rm x} -\tfrac{1}{4} \bar M {\rm x} +  f_{\rm aux} \ ,
\eeq
where
\begin{align}
&f \hspace{0.42cm} \ \equiv\  \tfrac{1}{3}(1+2\gamma)\, {\cal R}\ , 
\\
&f_{\mu \nu} \hspace{0.12cm}\ \equiv\   - 2(1 +2 \gamma)\, {\cal R}_{\mu \nu} \ ,\\
&f_{\rm aux} \ \equiv\   |\nabla_\mu F|^2  - \tfrac{1}{3} \gamma\, |F|^2\, {\cal R} \ .
\end{align}
The terms linear in $b_\mu$ couple to 
\beq
\displaystyle j^\mu \ \equiv\  - \tfrac{4i}{3}\left(  \nabla^\mu \phi \nabla^2 \phid - (\tfrac{1}{2}+\gamma) \nabla^\nu \phi \nabla_\nu \nabla^\mu \phid  +(1+\gamma) F \nabla^\mu \Fd  -  {\rm h.c.} \right)\ ,
\eeq
while the terms linear in $M$ couple to 
\beq
\displaystyle \bar {\rm x} \ \equiv\ \tfrac{4}{3}(1-2\gamma) \nabla_\mu F \nabla^\mu \phid + \tfrac{4}{3}(1-2\gamma) F \nabla^2 \phid \ .
\eeq
Therefore, ${\rm x} = 0$ if we take $\gamma= \tfrac{1}{2}$.  In this case, we find $f_{\mu \nu} = -4{\cal R}_{\mu \nu}$ and $f=\tfrac{2}{3} {\cal R}$, which indeed correspond to the conformal couplings, as desired.  We also find 
\beq
 j^\mu \ \equiv\  - \tfrac{4i}{3}\left(  \nabla^\mu \phi \nabla^2 \phid - \nabla^\nu \phi \nabla_\nu \nabla^\mu \phid  +\tfrac{3}{2} F \nabla^\mu \Fd  -  {\rm h.c.} \right)\ ,
\eeq
which matches the conformal FZ multiplet.
We have therefore demonstrated perfect consistency of our results with the expectations from linearized supergravity.

\section{Application: Supersymmetric Effective Theory of Inflation}
\label{sec:application}

A particularly interesting application of higher-derivative supergravity is to the {\it effective theory of inflation}~\cite{EFT1, EFT2}.  This theory describes the Goldstone boson associated with the spontaneous breaking of time translations during inflation. 
At lowest order in derivatives, the (universal) action describes fluctuations around slow-roll backgrounds.
However, (non-universal) higher-derivative terms are central to the predictions for higher-order correlation functions \cite{EFT1,Senatore:2009gt}. 
We will discuss the supersymmetric effective theory of inflation in detail in our companion paper~\cite{Paper2}. Here, we will derive a number of technical results that will be important for that work.

\vskip 4pt
In supersymmetric theories of inflation the (real) inflaton field $\varphi$ is promoted to a chiral superfield
\beq
\Phi = \phi + \cdots = \tfrac{1}{\sqrt{2}} (\sigma + i \varphi) + \cdots
\eeq
Notice that supersymmetry inevitably adds a second real scalar field $\sigma$. This additional scalar can have interesting implications for the phenomenology of inflation~\cite{Chen:2009zp,Paper2}.
To keep the inflaton naturally light, even in the presence of supergravity corrections, we assume an approximate shift symmetry for $\varphi$. 
This symmetry forbids superpotential couplings\footnote{It would be straightforward to include small shift symmetry breaking effects from a superpotential $W(\Phi)$, but for simplicity we focus our attention on the couplings related to the K\"ahler potential (\ref{equ:Kcs}). For the more general case, see \cite{Paper2}.} and restricts the K\"ahler potential to be a function of $\Phi + \Phid$, as well as arbitrary derivatives of $\Phi$.
The following derivative expansion of the K\"ahler potential is of particular interest~\cite{Paper2} (see also \cite{Justin})
\beq
\label{equ:Kcs}
K_{c_s} \ = \ \tfrac{1}{2}(\Phi+\Phid)^2 \left[ c_1 + c_2 \D_a \Phi \D^a \Phid + \cdots \right] \ \equiv \ c_1 K_1 + c_2 K_2 + \cdots
\eeq
The limit $c_2 \to 0$ corresponds to ordinary slow-roll inflation.
Turning on finite $c_2$ induces a speed of sound for the inflaton fluctuations $\delta \varphi$.
The physics of this theory is explained in much more detail in \cite{Paper2}. Here, we simply restrict our ambitions to the computation of the supergravity action:
\beq
\label{equ:Lcs}
 {\cal L}_{c_s} =   \int d^2 \Theta \, E\left[ -  \tfrac{1}{8}(\Db^2 - 8 R) K_{c_s} + \cdots \right] + {\rm h.c.}
\eeq

\subsection{Supersymmetric Slow-Roll Actions}

The lowest-order term in eq.~(\ref{equ:Kcs}), $K_1 \equiv \tfrac{1}{2}  (\Phi + \Phid)^2$, doesn't contain any derivatives, so the standard supergravity formulas for the scalar F-term potential \cite{WB} are applicable.
However, in order to make contact with the treatment for higher-derivative terms, it will be instructive not to integrate out the auxiliary fields $F$, $M$ and $b_\mu$.

From 
\beq
\mathcal{L}_1 \ =\  -  \tfrac{1}{16} \int d^2 \Theta\, E \left[  (\Db^2- 8R) (\Phi +\Phid)^2 + \cdots \right] + {\rm h.c.}\ ,
\eeq
we get the following expression for the scalar sector 
\begin{align}
{\cal L}_{1} &\ = \ \tfrac{1}{12} M \D^2 (\Phi +\Phid)^2|  + \tfrac{1}{64} \D^2 \Db^2 (\Phi +\Phid)^2|  +  \tfrac{1}{24} ( {\cal R} + \tfrac{2}{3}|M|^2 - \tfrac{2}{3}b^2) (\Phi +\Phid)^2| \ +\ {\rm h.c.}\ , \label{equ:L0}
\end{align}
where
\begin{align}
\D^2 (\Phi +\Phid)^2| &= 2 (\Phi +\Phid)| \D^2 \Phi| \ , \label{equ:a1}\\
\D^2 \Db^2 (\Phi +\Phid)^2| &= 2 \D^2 \Phi| \Db^2 \Phid| + 2 (\Phi +\Phid)| \D^2 \Db^2 \Phid| - 4 \D^\alpha \Db_{\dalpha} \Phid| \D_\alpha \Db^{\dalpha} \Phid| \ . \label{equ:a2}
\end{align}
To evaluate eqns.~(\ref{equ:a1}) and (\ref{equ:a2}), we use $- \tfrac{1}{4}\D^2\Phi| = F$ and $\D_\alpha \Db_{\dalpha} \Phid| = - 2i \sigma^\mu_{\alpha \dalpha} \partial_\mu \phid$. Moreover, repeated application of eq.~(\ref{equ:commutator}) gives
\begin{align}
\tfrac{1}{16} \D^2 \Db^2 \Phid| &= \Box \phid + \tfrac{2i}{3} b^\mu \partial_\mu \phid + \tfrac{2}{3} \bar M \bar F\ .
\end{align}
Hence, we find
\begin{align}
\tfrac{1}{12} M \,  \D^2 (\Phi + \Phid)^2|  +{\rm h.c.} &\ =\ - \tfrac{2}{3} (MF+\bar M \bar F) (\phi+\phid) \ , \, \label{equ:a1x}\\
\tfrac{1}{64}\D^2 \Db^2 (\Phi +\Phid)^2| +{\rm h.c.}&\ =\  \tfrac{1}{4} [\partial_\mu (\phi+\phid)]^2  + \tfrac{1}{4} [\partial_\mu (\phi-\phid)]^2 + \tfrac{1}{2}(\phi + \phid) \Box(\phi +\phid) \nonumber \\
&\hspace{1cm}+\, |F|^2 - \tfrac{1}{3}(\phi +\phid) \left[ i b^\mu \partial_\mu(\phi-\phid) - (MF+\bar M \bar F)\right]  \ . \label{equ:a2x}
\end{align}
Eq.~(\ref{equ:L0}) then becomes 
\begin{align}
{\cal L}^{(1)}_{\rm kin} & =   \tfrac{1}{4} [\partial_\mu (\phi+\phid)]^2  + \tfrac{1}{4} [\partial_\mu (\phi-\phid)]^2 + \tfrac{1}{2}(\phi + \phid) \Box(\phi +\phid)\ , \label{equ:Lkin}  \\
{\cal L}^{(1)}_{\rm aux} & = |F|^2 - \tfrac{1}{3} (\phi + \phid) \left[ i b^\mu \partial_\mu(\phi-\phid)+ (M F + \bar M \bar F) \right]  \label{equ:sig2}
\ , \\
{\cal L}_{\cal R}^{(1)} &=   \tfrac{1}{12} (\phi +\phid)^2 \left[ {\cal R} + \tfrac{2}{3}|M|^2 - \tfrac{2}{3}b^2 \right]\ .\label{equ:LR}
\end{align}
Integrating the last term in (\ref{equ:Lkin}) by parts, we get the standard kinetic terms
\begin{align}
{\cal L}^{(1)}_{\rm kin} &= -\tfrac{1}{2}(\partial_\mu \sigma)^2  -\tfrac{1}{2}(\partial_\mu \varphi)^2 \ .
\end{align}
The potential for the non-shift-symmetric field $\sigma$ is 
\begin{align}\label{equ:slowrollsigma}
{\cal L}_{\sigma}^{(1)} &\ =\ \tfrac{1}{6} ( {\cal R} + \tfrac{2}{3}|M|^2 - \tfrac{2}{3}b^2) \sigma^2  - \tfrac{\sqrt{2}}{3} (MF + \bar M \bar F)  \sigma + \tfrac{2}{3} b^\mu \partial_\mu \varphi\, \sigma  \ .
\end{align}
In de Sitter space, ${\cal R} = - 12 H^2$, we find a model-independent curvature-induced contribution to the mass of the partner of the inflaton, $\delta m_{\sigma}^2 = 4 H^2$.  
In addition, there will be other, more model-dependent contributions to the mass of $\sigma$.  Most importantly,  we need vacuum energy to have de Sitter space as a solution of Einstein's equations.  This vacuum energy may have nothing to do with $\Phi$ directly, and may be due to some additional spurion field $X$, whose F-term, ${\cal F}_X \neq 0$, breaks SUSY \cite{Paper2}.  Planck-suppressed couplings between these fields of the form 
\beq
\int d^4 \theta \, \frac{\beta}{12 \Mp^2} (\Phi + \Phid)^2 X^{\dagger} X\ ,
\eeq
will also contribute a mass of order $\beta H^2$.

The contributions from the auxiliary fields $M$ and $b_\mu$ to the potential for $\sigma$ are also model-dependent.  In the presence of a constant superpotential, $\langle W\rangle =W_0$, the field $M$ acquires a vev $\langle M\rangle = W_0 / \Mp^2$.  If $W_0 \sim H \Mp^2$, then $\langle M\rangle \sim H$ would also contribute significantly to the potential for $\sigma$.  In fact, in particle physics applications one often assumes theses large values for $W_0$ in order to cancel the cosmological constant.  However, in the context of inflation, the vacuum energy should not be cancelled. Moreover, there is no reason to prefer any particular value of $W_0$ and  it is therefore consistent to assume that the vev of $M$ is small.  Finally, a vev for $b_{\mu}$ spontaneously breaks spacetime translations and typically gives a negligible contribution to the potential. We refer interested readers to our companion paper \cite{Paper2}  for further details on these issues.

\subsection{Supersymmetric Small Speed of Sound}

The higher-derivative term in (\ref{equ:Kcs}), $K_2 = \tfrac{1}{2}  (\Phi + \Phid)^2\, \D_a \Phi \D^a \Phid$, is a bit more challenging and require the new results that we developed in this paper.

\vskip 10pt
\small
\hrule
\vskip 1pt
\hrule
\vskip 4pt
\small
\noindent
The Lagrangian can be written as
\begin{align}
{\cal L}_2 & \ =\  -\tfrac{1}{16} \int d^2 \Theta \, E\left[\, (\Db^2 - 8 R) (\Phi + \Phid)^2 \D_a \Phi \D^a \Phid \, \right] \ +\ {\rm h.c.}  \nonumber \\ 
&\ =\  \tfrac{1}{2} (\Phi +\Phid)^2|\, {\cal \tilde L}_{\partial \Phi \partial \Phid} \nonumber \\
& \hspace{0.7cm}+\, \tfrac{1}{12} M \,  \D^2 (\Phi + \Phid)^2| \D_a \Phi \D^a \Phid| \,+\, \tfrac{1}{64} \D^2 \Db^2 \big\{  (\Phi + \Phid)^2| \D_a \Phi \D^a \Phid|  \big\} \ +\ {\rm h.c.}\ , \label{equ:ted}
\end{align}
where ${\cal \tilde L}_{\partial \Phi \partial \Phid}$ is given by eq.~(\ref{equ:nobyparts}). 
We evaluate the second term in eq.~(\ref{equ:ted}) with the help of eq.~(\ref{equ:a1x}),
\beq
\tfrac{1}{12} M \,  \D^2 (\Phi + \Phid)^2| \D_a \Phi \D^a \Phid| +{\rm h.c.} = - \tfrac{2}{3} (MF+\bar M \bar F) (\phi+\phid) |\partial_\mu \phi|^2\ . \label{equ:no2}
\eeq
The last term in eq.~(\ref{equ:ted}) is a bit more involved
\begin{align}
&\tfrac{1}{64} \D^2 \Db^2 \big\{  (\Phi + \Phid)^2| \D_a \Phi \D^a \Phid|  \big\}  \, +\, {\rm h.c.} \ = \nonumber \\
&\hspace{0.7cm}=\ \tfrac{1}{64}\Big\{ \D^2 \Db^2 (\Phi + \Phid)^2| \D_a \Phi \D^a \Phid| + \D^2 (\Phi + \Phid)^2|  \Db^2( \D_a \Phi \D^a \Phid)| \nonumber \\
& \hspace{1.5cm}+\,  \Db^2 (\Phi + \Phid)^2|  \D^2( \D_a \Phi \D^a \Phid )|  - 4\, \D^\alpha \Db_{\dalpha} (\Phi + \Phid)^2 | \D_\alpha \Db^{\dalpha}(  \D_a \Phi \D^a \Phid)|  \Big\}  \, +\,  {\rm h.c.}  \label{equ:a3}
\end{align}
However, all the terms in (\ref{equ:a3}) have already been determined in previous calculations (see also Appendix~\ref{sec:toolkit}). Assembling the answer is therefore straightforward, if a bit tedious.
In fact, the first term in (\ref{equ:a3}) combines with (\ref{equ:no2}) into
\beq
\tfrac{1}{12} M \,  \D^2 (\Phi + \Phid)^2| \D_a \Phi \D^a \Phid|  + \tfrac{1}{64}   \D^2 \Db^2 (\Phi + \Phid)^2| \D_a \Phi \D^a \Phid| +{\rm h.c.} = \left[ {\cal L}_{\rm kin}^{(1)} + {\cal L}_{\rm aux}^{(1)} \right] |\partial_\mu \phi|^2 \ ,
\eeq
where ${\cal L}_{\rm kin}^{(1)}$ and ${\cal L}_{\rm aux}^{(1)}$ are given by (\ref{equ:Lkin}) and (\ref{equ:sig2}), respectively.
The next two terms in (\ref{equ:a3}) are related by complex conjugation, and therefore combine into
\begin{align} \tfrac{1}{32}  \Db^2 (\Phi + \Phid)^2| \D^2( \D_a \Phi \D^a \Phid)| \, +\, {\rm h.c.} &\ =\  ( \phi + \phid)\Big[  (F \partial_\mu\bar F \partial^\mu \phi + {\rm h.c.} ) - \tfrac{2i}{3} |F|^2  b_\mu \partial^\mu (\phi - \phid) \Big]\ .
\end{align}
The last term in (\ref{equ:a3}) gives
\begin{align}
&-\tfrac{1}{16} (\D^\alpha \Db_{\dalpha} (\Phi + \Phid)^2 |)(\D_\alpha \Db^{\dalpha}  \D_a \Phi \D^a \Phid|) \, +\, {\rm h.c.}  \ = \nonumber \\
&\hspace{1cm}=\ (\phi+\phid) \Big[\nabla^{\mu} \phi \nabla^\nu \phid \nabla_\nu \nabla_\mu \phid  - \tfrac{1}{6}M F (\partial^\mu \phid )^2 - \tfrac{1}{6} \bar M \bar F |\partial^\mu \phi|^2 \, +\, {\rm h.c.} \Big]  \ .
\end{align}
\hrule
\vskip 1pt
\hrule
\vskip 6pt
\normalsize 

\vskip 8pt
Combining all terms, we get  
\bea
\mathcal{L}_2 
&=&  \big[ {\cal L}_{\rm kin}^{(1)} + {\cal L}_{\rm aux}^{(1)} \big] |\partial_\mu \phi|^2 +  \tfrac{1}{2} \phi_+^2 \, \mathcal{\tilde L}_{\partial \phi \partial \phid}  \nonumber \\
&& +\, \tfrac{1}{2} \phi_+ \Big[ \partial_\mu |F|^2 \partial^\mu \phi_+ +  \left(F \partial_\mu\bar F - \bar F \partial_\mu F - \tfrac{4i}{3} |F|^2 b_\mu \right) \partial^\mu \phi_- \nonumber \\
&& \hspace{1.4cm}+\,
\tfrac{1}{2}\left(\partial^{\mu} \phi_+ \partial^\nu \phi_+ - \partial^{\mu} \phi_- \partial^\nu \phi_- \right) \nabla_\nu \nabla_\mu \phi_+  \nonumber \\
&& \hspace{1.4cm}+\, \tfrac{1}{6} (MF - \bar M \bar F) \partial_\mu \phi_+ \partial^\mu \phi_-  - \tfrac{1}{6}(MF+ \bar M \bar F) (\partial_\mu \phi_+)^2  \Big]\ , \label{equ:L2final}
\eea
where  we have defined the shorthand notation $\phi_\pm \equiv (\phi \pm \phid)$, i.e.~$\phi_+ = \sqrt{2}\sigma$ and $\phi_- = i \sqrt{2} \varphi$. Note that ${\cal L}_{\rm kin}^{(1)} $ and $ {\cal L}_{\rm aux}^{(1)} $ are given by eqs.~(\ref{equ:Lkin}) and (\ref{equ:sig2}), respectively, and $\mathcal{\tilde L}_{\partial \phi \partial \phid}$ was defined in eq.~(\ref{equ:nobyparts}).

\vskip 4pt
We will explore the physical implications of these results in our companion paper~\cite{Paper2}.
There we will find that the case of special interest for the supersymmetric effective theory of inflation is $\langle F \rangle = \langle M \rangle = \langle b_\mu \rangle \simeq 0$. The supergravity action then simplifies dramatically, 
\bea\label{eqn:csfinal}
{\cal L}_2 &=&  -\left(|\partial_\mu \phi|^2\right)^2  \, -\,   \partial_\mu \phi_+ \partial_\nu \phi_+ \partial^{\mu} \phi \partial^\nu \phid \nonumber \\
&& \, +\,  \tfrac{1}{2} \phi_+^2 \left[ \tilde {\cal L}_{\partial \phi \partial \phid} + \nabla_\mu \nabla_\nu \left(\partial^\mu \phi \partial^\nu \phid \right) + \tfrac{1}{2} \Box|\partial_\mu \phi|^2\right]   \ ,
\eea
where
\begin{align}
\tilde {\cal L}_{\partial \phi \partial \phid} &= \tfrac{1}{3} {\cal R} |\partial_\mu \phi|^2 -  {\cal R}_{\mu \nu} \partial^\mu \phi \partial^\nu \phid  +   \partial^\mu \phi \partial_\mu \Box \phid   \ .
\end{align}
In writing (\ref{eqn:csfinal}), we used $|\partial_\mu \phi|^2= \tfrac{1}{4}(\partial_\mu \phi_+)^2 - \tfrac{1}{4}(\partial_\mu \phi_-)^2$ and performed several integrations by part.  We see that curvature couplings in the higher-derivative part of the action can contribute significantly to the mass for the SUSY partner of the inflaton.
In fact, for the case of most interest for observations---i.e.~small sound speed, $c_s \ll 1$---the higher-derivative terms lead to a parametrically enhanced mass,
\beq
m_\sigma^2 \sim \frac{H^2}{c_s^2} \gg H^2 \ .
\eeq
Further discussion of this feature of small-$c_s$ supergravity will appear in \cite{Paper2}.

\newpage
\section{Conclusions}
\label{sec:conclusion}

Coupling higher-derivative effective theories to supergravity can be non-trivial.
This paper has developed the essential tools required for this task.
As an illustrative example, we applied the formalism to the effective theory of inflation~\cite{EFT1} (a theory of the Goldstone boson of spontaneously broken time translations during inflation).
This is an example in which the experimental predictions of various models depend sensitively on higher-derivative interactions. 
We computed the component action for a supersymmetric inflationary model in which a higher-derivative operator induces a propagation speed for the fluctuations that is different from the speed of light.  
We showed that curvature couplings associated with the higher-derivative terms lead to a parametrically enhanced mass for the scalar partner of the inflaton.  Details of the physical interpretation of our results have been relegated to a companion paper~\cite{Paper2}.

\vskip 4pt
Although this was beyond the scope of this toolkit paper, it would be interesting to apply the techniques that we developed here to other EFTs---in particular, other examples of spontaneous symmetry breaking.  In that case, the two-derivative action for a Goldstone boson is universal, and differences between models will only appear at higher orders in derivatives. Understanding the effects of higher-derivative operators then becomes particularly interesting.

\acknowledgments
We are grateful to Thomas~Dumitrescu, Guido~Festuccia, Vijay Kumar, and Nathan~Seiberg for helpful discussions.
D.B.~thanks the Institute for Advanced Study for hospitality while this work was being completed.
D.B.~gratefully acknowledges support from a Starting Grant of the European Research Council (ERC STG grant 279617) and partial support from STFC under grant ST/FOO2998/1. The research of D.G.~is supported by the DOE under grant number DE-FG02-90ER40542 and the Martin A.~and Helen Chooljian Membership at the Institute for Advanced Study.  

\newpage
\appendix
\section{SUGRA Toolbox}\label{sec:toolkit}

In this appendix, we will provide a list of identities that are necessary to reproduce the results in the main text.  
More identities can be found in Wess and Bagger~\cite{WB}.

\subsection{Collection of Useful Identities}

The most important identity of the entire paper is
\beq
\label{equ:commutatorA}
\left(\D_{C} \D_B - (-1)^{bc} \D_B \D_C \right) V^{A} = (-1)^{d(c+b)} V^D  R_{CB D}{}^A - T_{CB}^D \D_D V^A \ ,
\eeq
where $b$, $c$ and $d$ are functions of $B$, $C$ and $D$, respectively. These functions take the values zero or one, depending on whether $B$, $C$ and $D$ are vector or spinor indices.
The contractions in eq.~(\ref{equ:commutatorA}) are to be understood as follows 
\beq
V^A W_A = V^{a} W_{a} + V^\alpha W_\alpha + V_{\dalpha} W^{\dalpha}\ .
\eeq
Note that the raised and lowered indices have been flipped in the last term.

In minimal supergravity, the non-vanishing components of the torsion are given by
\begin{align}
T_{\alpha \dot \alpha}^{a} &\ =\ 2i \sigma_{\alpha \dot \alpha}^a \ , \\
T_{\dot \beta a}^{\alpha}&\ =\ - i \sigma_{\dot \beta a}^{\alpha} R\ , \label{equ:T2} \\
T_{\beta a}^{\alpha} &\ =\ \tfrac{i}{8} \bar\sigma_a^{\gamma \dot\gamma}(\delta_{\gamma}^{\alpha} G_{\beta \dot\gamma} - 3 \delta_{\beta}^{\alpha} G_{\gamma \dot \gamma} + 3 \epsilon_{\beta \gamma} G^{\alpha}_{\dot\gamma})\ , \label{equ:T3}
\end{align}
where $G_{\alpha \dalpha}$ is real and $R$ is chiral.  Expressions with dotted and undotted indices interchanged (and $R \to \Rd$) are obtained by complex conjugation.  All other components of the torsion vanish.
A number of useful facts, related to the components in (\ref{equ:T2}) and (\ref{equ:T3}), follow directly from the Bianchi identities:
\begin{align}
\bar\D_{\dot\alpha} R &\ =\ 0\ , \\
\D^{\alpha} G_{\alpha \dot \alpha} &\ =\ \bar \D_{\dot \alpha} \Rd \ , \label{equ:16}\\
R_{\alpha \beta \gamma \delta} &\ =\ 4 (\epsilon_{\alpha \gamma} \epsilon_{\beta \delta}+\epsilon_{\alpha \delta} \epsilon_{\beta \gamma} ) \Rd \ , 
\label{equ:18}\\
R_{\alpha \beta \dot \gamma \dot \delta} &\ =\ 0 \ .
\label{equ:19} 
\end{align}
These relations are useful for solving for the components of the superfield $G_a \equiv - \tfrac{1}{2} \bar \sigma^{\alpha \dot \alpha}_a G_{\alpha \dot \alpha}$ (see \S\ref{sec:proofG}). 

In determining the contributions to the action in components, we will frequently require the various $\Theta$-components of $R$ and $G_a$.  Because $R$ is chiral, its components are more easily determined and are given by
\beq
R = - \tfrac{1}{6} \left\{ M + \Theta^2 \left(-\tfrac{1}{2}{\cal R} + \tfrac{2}{3}|M|^2 + \tfrac{1}{3} b_a b^a - i\D_a b^a \right) \right\} \ +\ {\rm fermions}\ , \label{equ:Rcom}
\eeq
where ${\cal R}$ is the Ricci scalar and $b_a \equiv - 3 G_a|$.
The components of $G_a$ are more difficult to determine.  Since in this paper we are mostly interested in scalar couplings, we will not keep track of fermions. The complete list of components of $G_a$ can then be written as
\begin{align}
G_{a} | &\ =\ - \tfrac{1}{3} b_{a}\ , \label{eqn:glowest} \\
\D_{\alpha} G_{a} | &\ =\  0 \ ,\\
\bar \D_{\dot \alpha} G_{a} | &\ =\ 0\ , \\
- \tfrac{1}{4}\D^2 G_{a} | &\ =\   + \tfrac{i}{6} (\D_{a} +i b_a) \Md\ , \label{equ:A14}\\
-\tfrac{1}{4} \bar\D^2 G_{a} | &\ =\  - \tfrac{i}{6} (\D_{a}-ib_a) M\ , \label{equ:A15} \\
\bar \D_{\dot \alpha}\D^2 G_{a} | &\ =\ 0 \ ,\\
\D_{\alpha} \bar\D^2 G_{a} | &\ =\ 0  \ ,
\end{align}
and
\bea
\bar \sigma^{\alpha \dot \alpha}_b \D_{\alpha}  \bar \D_{\dot \alpha}G_a|  &=&  \left(  \tfrac{1}{6}{\cal R} + \tfrac{1}{9}|M|^2 + \tfrac{1}{9} b^2 \right) \eta_{ab}- {\cal R}_{ab} - \tfrac{2 i}{3} \D_b b_a-\epsilon^{cd} {}_{ab} \D_c G_d + \tfrac{2}{9} b_a b_b \ , \label{equ:A18}\\
\tfrac{1}{16} \D^2 \bar\D^2 G_a| &=& i  \left(\D_{a} -\tfrac{i}{3}b_a \right)  \left( -\tfrac{1}{4}\D^2 R| \right)  + \tfrac{i}{36} \Md \D_a M + \tfrac{i}{36} M  \D_{a} \bar M - \tfrac{1}{12} b_a |M|^2 \label{eqn:ghighest} \ .
\eea
We will present derivations of these results in \S\ref{sec:proofG}.

\subsection{Collection of One-Derivative Results}

Here, we collect results for the bosonic components of $\D_{a} \Phi$ and $\D_a \Phid$.
These results serve as key building blocks for deriving the supergravity couplings of general higher-derivative theories.

\subsubsection*{Components of \ `\ $\D_a \Phi$\ '}

Showing the bosonic terms only, we get
\begin{align}
\D_a \Phi | &= e^{\mu}_a\, \nabla_{\mu} \phi \ , \\
- \tfrac{1}{4} \Db^2 \D_a \Phi| &=  e^{\mu}_a \left[- \tfrac{1}{6} M  \nabla_{\mu} \phi \right]\ , \\
- \tfrac{1}{4} \D^2 \D_a \Phi| &=  e^{\mu}_a \left[ \tfrac{1}{6} \bar M  \nabla_{\mu} \phi+ (\nabla_\mu + \tfrac{2i}{3} b_\mu) F  \right]\ , \\
\tfrac{1}{16} \D^2 \Db^2 \D_a \Phi| &= e^{\mu}_a \left[ (\tfrac{1}{6}{\cal R} - \tfrac{5}{18}|M|^2 - \tfrac{1}{9} b^2 +  \tfrac{i}{3}\nabla \hskip -3pt \cdot \hskip -2pt b) \nabla_\mu \phi - \tfrac{1}{6} M \nabla_\mu F - \tfrac{i}{9} b_\mu MF\right] \ .
\end{align}
At times we also need the following mixed derivatives
\beq
\D^\alpha \Db^{\dalpha} \D_a \Phi| = e^\mu_a\, \tfrac{i}{3} M F \bar \sigma_{\mu}^{\alpha \dalpha}\ .
\eeq

\subsubsection*{Components of \ `\ $\D_a \Phid$\ '}

Many of the components of $\D_a \Phid$ can be related to the components of $\D_a \Phi$ by complex conjugation:
\begin{align}
\D_a \Phid | &= e^{\mu}_a\, \nabla_{\mu} \phid \ , \\
- \tfrac{1}{4} \D^2 \D_a \Phid| &=  e^{\mu}_a \left[- \tfrac{1}{6} \bar M  \nabla_{\mu} \phid \right]\ , \\
- \tfrac{1}{4} \Db^2 \D_a \Phid| &=  e^{\mu}_a \left[ \tfrac{1}{6}  M  \nabla_{\mu} \phid + (\nabla_\mu - \tfrac{2i}{3} b_\mu) \bar F  \right]\ . 
\end{align}
The computation of the term $\D^2 \Db^2 \D_a \Phid|$, on the other hand, requires a significant amount of work (see \S\ref{sec:business}). Ultimately, we find
\begin{align}
\tfrac{1}{16} \D^2 \Db^2 \D_a \Phid| &= e^{\mu}_a \left[\nabla_\mu \nabla^2 \phid + \tfrac{2i}{3} b^\nu \nabla_\mu \nabla_\nu \phid + \tfrac{1}{2} F_\mu{}^\nu \nabla_\nu \phid + F_\mu \right] \ ,
\end{align}
where
\bea
F_{\mu} &\equiv&  \tfrac{5}{6} \Md  \nabla_\mu \Fd + \tfrac{1}{3} \Fd \nabla_\mu \Md - \tfrac{i}{3} b_\mu \Md \Fd\ , \\
F_\mu{}^\nu &\equiv& \left(\tfrac{1}{6} {\cal R} + \tfrac{1}{6} |M|^2 + \tfrac{1}{3} b^2 - \tfrac{i}{3} \nabla \hskip -3pt \cdot \hskip -2pt b \right) \delta^\nu_\mu  \nonumber \\
&& +  \left( -2{\cal R}_{\mu}{}^{\nu} - \tfrac{4 i}{3} \nabla^\nu b_\mu +\tfrac{4i}{3} \nabla_\mu b^\nu - \tfrac{4}{9}  b_\mu b^\nu  + \tfrac{2}{3} \epsilon^{\rho \sigma}{}_\mu{}^\nu \nabla_\rho b_\sigma \right) \, . 
\eea
At times we also need the following mixed derivatives
\beq
\D^\alpha \Db^{\dalpha} \D_a \Phid| = e^\mu_a \left[ \tfrac{i}{3} \bar M \bar F \bar \sigma_{\mu}^{\alpha \dalpha} - 2i \sigma_\nu^{\alpha \dalpha} \nabla^\nu \nabla_\mu \phid - \tfrac{2i}{3} \epsilon_{\mu \nu}{}^{\tau \rho} \bar \sigma^{\alpha \dalpha}_{\rho}\, b_\tau \nabla^\nu \phid \, \right]\ .
\eeq

\section{Derivations of Selected SUGRA Identities} \label{sec:proofs}

In this appendix, we will derive a few of the identities from Appendix~\ref{sec:toolkit}.
This will illustrate how to derive all of the results presented in this paper.

\subsection{Components of \,$G_a$}\label{sec:proofG}

When dealing with higher-derivative theories, computing the action in components requires knowledge of all components of $G_a$.  While these components all follow from the solutions to the Bianchi identities, deriving the formulas is not always easy.  The results were given in eqs.~(\ref{eqn:glowest})--(\ref{eqn:ghighest}).  Here, we will derive a few of these results to show how this is done (and since not all these formulas appear in standard references like \cite{WB}).

\vskip 6pt
Most components can be derived simply from eq.~(\ref{equ:16}), i.e.~$\D^{\alpha} G_{\alpha \dalpha} = \Db_{\dalpha} \bar R$.  
For example, to derive $\D^2 G_{\alpha \dot \alpha}$, we perform the following manipulations
\bea
2 \D_{\alpha} \D_{\beta} G_{\gamma \dot \gamma} &=& (\D_{\alpha} \D_{\beta} + \D_{\beta} \D_{\alpha}) G_{\gamma \dot \gamma} + (\D_{\alpha} \D_{\beta} - \D_{\beta} \D_{\alpha}) G_{\gamma \dot \gamma} \nonumber \\
& = &  ( -\tfrac{1}{2}R_{\alpha \beta \delta \dot \delta \gamma \dot \gamma} G^{\delta \dot \delta}  - T_{\alpha \beta}^{D} \D_D G_{\gamma \dot \gamma} ) + \epsilon_{\alpha \beta} \D^2 G_{\gamma \dot \gamma} \nonumber \\
& = & - \epsilon_{\dot \delta \dot \gamma} R_{\alpha \beta \delta \gamma} G^{ \delta \dot \delta}  + \epsilon_{\alpha \beta} \D^2 G_{\gamma \dot \gamma} \nonumber \\
& = &4 \Rd (  \epsilon_{\beta \gamma}  G_{\alpha \dot \gamma}+  \epsilon_{\alpha \gamma} G_{\beta \dot \gamma}) +   \epsilon_{\alpha \beta} \D^2 G_{\gamma \dot \gamma}\ , \label{equ:114}
\eea
where we used $T^D_{\alpha \beta} = 0$ and $ R_{N M \delta \dot \delta \gamma  \dot \gamma} = -2  \epsilon_{\delta \gamma} R_{N M \dot \delta \dot \gamma} + 2 \epsilon_{\dot \delta \dot \gamma} R_{N M \delta \gamma}$, together with (\ref{equ:18}) and (\ref{equ:19}).  Contracting both sides with $\epsilon^{\gamma \beta}$ and using (\ref{equ:16}), we find
\begin{align}
2 \D_{\alpha} \D^{\gamma} G_{\gamma \dot \gamma} \ =\ 2 \D_{\alpha} \bar\D_{\dot \gamma} \Rd &\ =\  12 \Rd 
\, G_{\alpha \dot \gamma} -  \D^2 G_{\alpha \dot \gamma} \ ,\end{align}
or 
\beq
\label{equ:DDG1}
  \D^2 G_a \ =\  4i(\D_a-3i G_a) \Rd \ , 
  \eeq
where $G_a \equiv - \frac{1}{2} \bar\sigma^{\alpha \dot \gamma}_a G_{\alpha \dot \gamma}$. Taking the lowest component of (\ref{equ:DDG1}) reproduces eq.~(\ref{equ:A14}). Eq.~(\ref{equ:A15}) is related to this by complex conjugation. Moreover, we get eq.~(\ref{eqn:ghighest}) from the identity
\bea
\D^2 \bar\D^2 G_a| &=&  -4i \big[\D_a + i G_a|  \big] \D^2 R| +16 i \bar R| \D_a R| +12 R| \, \D^2 G_a|\ ,
\eea
where we used  $T^{\beta}_{\beta a} =2 i G_a$ and $T^{\dbeta}_{\beta a} =  i \sigma^{\dbeta}_{\beta a} \Rd$ to determine $\D^2 \D_a R| =  (\D_a - 2iG_a|) \D^2 R| + 4  \Rd| \D_a R|$.
Substituting the components of $R$ from eq.~(\ref{equ:Rcom}), leads to eq.~(\ref{eqn:ghighest}).

Deriving the mixed components of $G_a$, eq.~(\ref{equ:A18}), requires a bit more work. However, conceptually the computation is straigthforward, so we only sketch the derivation.
We start with the component $\bar \D_{\dot \beta} \D_{\beta} G_{\alpha \dot \alpha}$.  First, we use $\bar \D_{\dot \beta} \D^{\alpha} G_{\alpha \dot \alpha} = - \frac{1}{2} \epsilon_{\dot \beta \dot \alpha} \bar \D^2 \Rd$ to determine that 
\bea
\bar \D_{\dot \beta} \D_{\beta} G_{\alpha \dot \alpha} &=& - \tfrac{1}{4} \epsilon_{ \beta \alpha} \epsilon_{\dot \beta \dot \alpha} \bar \D^2 \Rd-\tfrac{1}{4}\epsilon_{\dbeta \dalpha} (\Db_{\dgamma} \D_{\beta} G^{\dgamma}_{\alpha} + \Db_{\dgamma} \D_{\alpha} G^{\dgamma}_{\beta} )  + \tfrac{1}{4}( \bar\D_{\dot \beta} \D_{\beta} G_{\alpha \dot \alpha} + {\rm perms.} ) \nonumber \\
&=& - \tfrac{1}{4} \epsilon_{ \beta \alpha} \epsilon_{\dot \beta \dot \alpha} \bar \D^2 \Rd+\tfrac{i}{2}\epsilon_{\dbeta \dalpha} (\D_{\dgamma \beta} G^{\dgamma}_{\alpha} + \D_{\dgamma \alpha} G^{\dgamma}_{\beta} )  + \tfrac{1}{4}( \bar\D_{\dot \beta} \D_{\beta} G_{\alpha \dot \alpha} + {\rm perms.} ) \ , \label{equ:118} 
\eea
where $\D_{\alpha \dalpha} \equiv \sigma_{\alpha \dalpha}^a \D_a$ and the last term is symmetric under $\alpha \leftrightarrow \beta$ and ${\dot \alpha} \leftrightarrow \dot \beta $.  This symmetric combination of mixed derivatives of $G_{\alpha \dot \alpha}$ can be determined from the solutions to the Bianchi identities. Specifically, we use~\cite{WB} 
\bea
 (\sigma^{ab} \epsilon)_{ \beta \alpha}(\epsilon \bar \sigma^{cd})_{ \dot \beta \dot \alpha} R_{abcd}  &=& 4 (G_{\beta \dot \alpha} G_{\alpha \dot\beta} + G_{\beta \dot \beta} G_{\alpha \dot \alpha} ) + 2 i (\D_{\beta \dot\beta} G_{\alpha \dot \alpha}+ {\rm perms.}) \nonumber \\ && +\, 2 ( \bar\D_{\dot \beta} \D_{\beta} G_{\alpha \dot \alpha} + {\rm perms.} )\ , \label{equ:120}
\eea
where $\sigma^{ab} \equiv \tfrac{1}{4}\big( \sigma^a \bar \sigma^{b} - \sigma^b \bar \sigma^{a}\big)$ and $\bar \sigma^{ab} \equiv \tfrac{1}{4}\big( \bar \sigma^{ a }  \sigma^{b} - \bar \sigma^{ b}  \sigma^{a} \big)$.
Contracting this with $\bar \sigma^{\alpha \dalpha}_{a} \bar \sigma^{\beta \dot \beta}_b$, using various identities for traces of $\sigma$-matrices (cf.~Appendix A of \cite{WB}), and symmetries of the Riemann tensor, we find 
\bea
\bar \sigma^{\beta \dot \beta}_b \bar \D_{\dot \beta} \D_{\beta} G_a  &=&  \left[ - \tfrac{1}{6}{\cal R} - \tfrac{1}{9}|M|^2 -G_c G^c \right] \eta_{ab}+ {\cal R}_{ab} \nonumber\\ 
&&+\, 2 i \D_b G_a+\epsilon^{cd} {}_{ab} \D_c G_d+ 2 G_a G_b\ .
\eea
Substituting this into the identity
\beq
\bar \sigma^{\beta \dot \beta}_b \D_{\beta}  \bar \D_{\dot \beta}G_a 
 =  4 i \, \D_b G_a - \bar \sigma^{\beta \dot \beta}_b \Db_{\dbeta} \D_\beta G_a\ , \label{equ:DDG2}
\eeq
reproduces eq.~(\ref{equ:A18}).  We could have derived the same result using complex conjugation with $[\D_{\beta} \Db_{\dbeta} G_{a}]^{\dagger} = - \Db_{\dbeta} \D_{\beta} G_a$.

\subsection{Derivation of \,$\D^2 \Db^2 \D_a \Phid|$}
\label{sec:business}

The supergravity treatment of higher-derivative terms requires us to compute the components of $\D_a \Phi$ and $\D_a \Phid$. We do this by acting with spinor derivatives $\D_{\alpha}$ and $\Db_{\dalpha}$ and then taking lowest components.
By far the most involved computation is $\D^2 \Db^2 \D_a \Phid|$. Hence, rather than showing all of our computations, we will only present the derivation of this term.
After the reader understands this computation, she should have no problem to reproduce all other results in this paper.

\vskip 6pt
Repeated application of eq.~(\ref{equ:commutatorA}) leads to 
\bea
\D^2 \Db^2 \D_a \Phid 
&=& \D_a \D^2 \Db^2 \Phid - \epsilon^{\beta \alpha} T_{\alpha a}^D \D_D \D_\beta \Db^2 \Phid -\epsilon^{\beta \alpha} R_{\alpha a \gamma \beta} \D^{\gamma} \Db^2 \Phid  - \epsilon^{\beta \alpha} \D_{\alpha} (T_{\beta a}^D \D_D \Db^2 \Phid) \nonumber \\
&& -\  \epsilon^{\dalpha \dbeta} \D^2( T_{\dalpha a}^{D} \D_D \Db_{\dbeta}\Phid) -  \epsilon^{\dalpha \dbeta} \D^2 (R_{\dalpha a \dot \gamma \dbeta} \Db^{\dot \gamma} \Phid) -  \epsilon^{\dalpha \dbeta} \D^2\Db_{\dalpha}(T_{\dbeta a \dot \gamma}  \Db^{\dot \gamma} \Phid) \ .  \label{equ:business}
\eea
The hard part is to expand each of these contributions in terms of the fundamental scalar components.
To make this manageable, we are going to start dropping fermions. We will discuss the terms of eq.~(\ref{equ:business}) one by one:

\vskip 4pt
- The bottom component of the first term is
\bea
 (1) \equiv \tfrac{1}{16} \D_a \D^2 \Db^2 \Phid|  &=&  e^{\mu}_a\, \nabla_\mu \left( \hat \Box \phid  + \tfrac{2}{3} \Md \Fd \right)\ ,
\eea
where $\hat \Box \phid \equiv (\nabla_\nu +\tfrac{2i}{3} b_\nu) \nabla^\nu \phid$.

\vskip 4pt
- The second term can be manipulated as follows
\bea
(2) \equiv \tfrac{1}{16} \epsilon^{\beta \alpha} T_{\alpha a}^D \D_D \D_\beta \Db^2 \Phid &=& \tfrac{1}{16}\epsilon^{\beta \alpha} T_{\alpha a}^\gamma \D_\gamma \D_\beta \Db^2 \Phid + \tfrac{1}{16} \epsilon^{\beta \alpha} T_{\alpha a \dot \gamma} \Db^{\dot \gamma} \D_\beta \Db^2 \Phid \nonumber \\
 &=& \tfrac{1}{32} T^{\alpha}_{\alpha a} \D^2 \Db^2 \Phid + \tfrac{1}{16} \epsilon^{\beta \alpha} T_{\alpha a \dot \gamma} \Db^{\dot \gamma} \D_\beta \Db^2 \Phid \ ,
 \eea
 where 
 \bea
 \tfrac{1}{32} T^{\alpha}_{\alpha a} \D^2 \Db^2 \Phid | & = & e^\mu_a \left[ - \tfrac{i}{3}b_\mu \left( \hat \Box \phid + \tfrac{2}{3}\Md \Fd \right) \right]\ , \eea
 and
 \bea
\tfrac{1}{16}  \epsilon^{\beta \alpha} T_{\alpha a \dot \gamma} \Db^{\dot \gamma} \D_\beta \Db^2 \Phid &=& \tfrac{1}{16}  \epsilon^{\beta \alpha} T_{\alpha a \dot \gamma} \left(-2i \sigma^{\dot \gamma b}_{ \beta} \D_b \Db^2 \Phid - \D_\beta \Db^{\dot \gamma}\Db^2 \Phid \right) \nonumber \\
 &=& \tfrac{1}{16}  \epsilon^{\beta \alpha} T_{\alpha a \dot \gamma} \left(-2i \sigma^{\dot \gamma b}_{ \beta} \D_b \Db^2 \Phid - \D_\beta (8 R) \Db^{\dot \gamma} \Phid \right) \nonumber \\
 & = & \tfrac{1}{16}  \epsilon^{\beta \alpha} T_{\alpha a \dot \gamma} \left(-2i \sigma^{\dot \gamma b}_{ \beta} \D_b \Db^2 \Phid +2i   (8 R) \sigma^{\dot \gamma b}_{ \beta} \D_b \Phid \right) + \cdots\nonumber \\
  & = &  e^\mu_a \left[ - \tfrac{1}{6} \Md \nabla_\mu \Fd + \tfrac{1}{18} |M|^2 \nabla_\mu \phid \, \right] + \cdots
 \eea
Hence, we find
\bea
(2) &=& e^\mu_a \left[- \tfrac{i}{3}  b_\mu \Big( \hat \Box \phid + \tfrac{2}{3} \Md \Fd \Big)  -  \tfrac{1}{6} \Md \nabla_\mu \Fd + \tfrac{1}{18} |M|^2 \nabla_\mu \phid\,  \right] \ . \ \ \ \ \
\eea

\vskip 4pt
- The third term is proportional to fermions
\bea
(3) \equiv \epsilon^{\beta \alpha} R_{\alpha a \gamma \beta} \D^{\gamma} \Db^2 \Phid &=& 0 + \cdots
\eea

\vskip 4pt
- The fourth term can be written as
\bea
(4) \equiv \tfrac{1}{16} \epsilon^{\beta \alpha} \D_{\alpha} (T_{\beta a}^D \D_D \Db^2 \Phid)& = &  \tfrac{1}{16}\epsilon^{\beta \alpha}T_{\beta a}^D \D_{\alpha}  \D_D \Db^2 \Phid \nonumber \\
& =& \tfrac{1}{32} T_{\alpha a }^{\alpha} \D^2 \Db^2 \Phid +  \tfrac{1}{16} \epsilon^{\beta \alpha}T_{\beta a \dot \delta} \D_{\alpha}  \Db^{\dot \delta} \Db^2 \Phid \nonumber \\
& =& \tfrac{1}{32} T_{\alpha a }^{\alpha} \D^2 \Db^2 \Phid +  \tfrac{1}{16}\epsilon^{\beta \alpha}T_{\beta a \dot \gamma} \D_{\alpha}  (8 R) \Db^{\dot \gamma} \Phid \nonumber \\
&=& \tfrac{1}{32} T_{\alpha a }^{\alpha} \D^2 \Db^2 \Phid - i \epsilon^{\beta \alpha}T_{\beta a \dot \gamma} \, R\, \sigma^{\dot \gamma b}_{\alpha} \D_b \Phid \ .
\eea
We therefore find 
\bea
(4) &=&  e^\mu_a \left( - \tfrac{i}{3} b_\mu \Big( \hat \Box \phid + \tfrac{2}{3}\Md \Fd \Big) + \tfrac{1}{18} |M|^2 \nabla_\mu \phid \right)\ .
\eea

\vskip 4pt
- The fifth term is
\bea
(5) \equiv  \tfrac{1}{16} \epsilon^{\dalpha \dbeta} \D^2 ( T_{\dalpha a}^{D} \D_D \Db_{\dbeta}\Phid) &=& - \tfrac{1}{32} \D^2(T_{\dalpha a}^{\dalpha} \Db^2 \Phid)  - \tfrac{i}{8}  \epsilon^{\dalpha \dbeta} \sigma_{\gamma \dbeta}^b \D^2 (T_{\dalpha a}^{\gamma}  \D_b\Phid)\ , 
 \eea
where 
\bea
-\tfrac{1}{32} \D^2(T_{\dalpha a}^{\dalpha} \Db^2 \Phid)| &= &
- \tfrac{1}{32} T^{\dot \alpha}_{\dot \alpha a}|  \D^2  \bar \D^2 \Phid| -\tfrac{1}{32} \D^2 T^{\dot \alpha}_{\dot \alpha a}| \bar \D^2 \Phid| \nonumber \\ 
&=&
- \tfrac{i}{16} G_a| \D^2 \bar \D^2 \Phid| -\tfrac{i}{16} \D^2 G_a| \bar \D^2 \Phid|  \nonumber \\
&=&   e^\mu_a \left[    \tfrac{i}{3}  b_\mu  \Big(\hat  \Box \phid + \tfrac{2}{3}\Md \Fd \Big)  + \tfrac{1}{6}  \Fd (\nabla_{\mu} +i b_\mu) \Md \right] \ ,
\eea
and 
\bea
-\tfrac{i}{8}  \epsilon^{\dalpha \dbeta} \sigma_{\gamma \dbeta}^b \D^2 (T_{\dalpha a}^{\gamma}  \D_b\Phid)  &=& - \tfrac{1}{4} \D^2 (R \, \D_a \Phid )\nonumber \\ 
&=& - \tfrac{1}{4} \D^2 R| \D_a \Phid| - \tfrac{1}{4} R| \D^2 \D_a \Phid| \nonumber\\
&=& \left(- \tfrac{1}{4} \D^2 R| +  R| \bar R| \right) \D_a \Phid| \nonumber\\
&=&  e^\mu_a   \left( - \tfrac{1}{4} \D^2 R| + \tfrac{1}{36} |M|^2  \right) \nabla_\mu \phid \,  \ .
\eea
Putting it all together, we find
\bea
(5) &=&
 e^\mu_a \left[  \tfrac{i}{3}  b_\mu  \Big(\hat  \Box \phid + \tfrac{2}{3}\Md \Fd \Big)   + \tfrac{1}{6}  \Fd (\nabla_{\mu} +i b_\mu) \Md  + \left( - \tfrac{1}{4} \D^2 R| + \tfrac{1}{36} |M|^2  \right) \nabla_\mu \phid  \right] \ . 
\eea

\vskip 6pt
The sum of the first five terms is
\bea
({\rm I}) \ \equiv \ (1)-(2)-(3)-(4)-(5) &=& e^a_\mu \Big[ (\nabla_\mu + \tfrac{i}{3} b_\mu) \bigl(\hat \Box \phid + \tfrac{2}{3} \Md \Fd \bigr) + \left(\tfrac{1}{4} \D^2 R| -\tfrac{5}{36} |M|^2 \right) \nabla_\mu \phid  \nonumber \\
&& \hspace{1cm} -\ \tfrac{i}{6} b_\mu \Md \Fd + \tfrac{1}{6} \Md \nabla_\mu \Fd - \tfrac{1}{6} \Fd \nabla_\mu \Md \Big]\ .
\eea

\vskip 6pt
The computation of the last two terms in eq.~(\ref{equ:business}) is a bit more involved:

\vskip 4pt
- The sixth term is
\bea
(6) \equiv \tfrac{1}{16} \epsilon^{\dalpha \dbeta} \D^2 (R_{\dalpha a \dot \gamma \dbeta} \Db^{\dot \gamma} \Phid) &=&  \tfrac{1}{16} \epsilon^{\dalpha \dbeta} \D^\alpha \left[ - R_{\dalpha a \dot \gamma \dbeta}  \D_\alpha   \Db^{\dot \gamma} \Phid + \D_\alpha R_{\dalpha a \dot \gamma \dbeta} \Db^{\dot \gamma} \Phid  \right] \\
&=&   \tfrac{1}{16} \epsilon^{\dalpha \dbeta} \left[ R_{\dalpha a \dot \gamma \dbeta}  \D^2  \Db^{\dot \gamma} \Phid - 2 \D^\alpha R_{\dalpha a \dot \gamma \dbeta}  \D_\alpha \Db^{\dot \gamma} \Phid + \Db^{\dot \gamma} \Phid \D^2 R_{\dalpha a \dot \gamma \dbeta} \right] \ . \nonumber
\eea
Since only the middle term on the r.h.s. is non-fermionic, we find
\bea
(6) &=&  \tfrac{i}{8}   \epsilon^{\dalpha \dbeta}\, \bar \sigma^{\delta \ddelta}_a\, \D_\alpha R_{\dalpha \delta \ddelta \dgamma \dbeta}|\,  \bar \sigma^{\alpha \dgamma}_{b}\D^b \phid\ .
\eea
The evaluation of  $\D_\alpha R_{\dalpha \delta \ddelta \dgamma \dbeta}|$ involves mixed derivatives of $G_{\alpha \dalpha}$. Determining those requires quite a bit of extra work.
First, we note that
\bea
\epsilon^{\dalpha \dbeta} R_{\dalpha \delta \dot \delta \dot \gamma \dbeta} &=& -\, 3i \epsilon_{\dot \delta \dot \gamma} \Db_{\dot \epsilon} G^{\dot \epsilon}_\delta - \tfrac{i}{2}(\Db_{\dot \gamma} G_{\delta \dot \delta} + \Db_{\dot \delta} G_{\delta \dot \gamma}) -\tfrac{3 i}{2} \Db_{\dot \delta} G_{\delta \dot \gamma} \nonumber \\
&=& -\, \tfrac{9i}{4} \epsilon_{\dot \delta \dot \gamma} \Db_{\dot \epsilon} G^{\dot \epsilon}_\delta - \tfrac{5i}{4}(\Db_{\dot \gamma} G_{\delta \dot \delta} + \Db_{\dot \delta}G_{\delta \dot \gamma}) \nonumber \\
&=&  +\, \tfrac{9i}{4} \epsilon_{\dot \delta \dot \gamma} \D_{\delta} R - \tfrac{5i}{4}(\Db_{\dot \gamma} G_{\delta \dot \delta} + \Db_{\dot \delta}G_{\delta \dot \gamma})\ ,
\eea
where the first line uses a solution of the Bianchi identities \cite{WB} to relate $R_{\dalpha \delta \ddelta \dgamma \dbeta}$ to derivatives of $G_{\alpha \dalpha}$, and the second line uses the identity
\beq
\Db_{\ddelta} G_{\delta \dgamma} - \Db_{\dgamma} G_{\delta \ddelta} = - \epsilon_{\ddelta \dgamma} \Db_{\dot \epsilon} G^{\dot \epsilon}_\delta\ .
\eeq
Hence, we obtain
\bea
\epsilon^{\dalpha \dbeta} \D_\alpha R_{\dalpha \delta \dot \delta \dgamma \dbeta}| &=& \tfrac{9i}{4} \epsilon_{\dot \delta \dot \gamma} \D_{\alpha} \D_{\delta} R| - \tfrac{5i}{4}\D_{\alpha}(\Db_{\dot \gamma} G_{\delta \dot \delta} + \Db_{\dot \delta}G_{\delta \dot \gamma})| \\
&=&  \tfrac{9i}{8} \epsilon_{\dot \delta \dot \gamma} \epsilon_{\alpha \delta} \D^2 R| - \tfrac{5i}{8}(\D_{\alpha} \Db_{\dot \gamma} G_{\delta \dot \delta} + {\rm perms.})| - \tfrac{5i}{8} \epsilon_{\alpha \delta} \D^{\beta}(\Db_{\dot \gamma} G_{\beta \dot \delta} + \Db_{\dot \delta}G_{\beta \dot \gamma})|\nonumber \\
&=&   \tfrac{9i}{8} \epsilon_{\dot \delta \dot \gamma} \epsilon_{\alpha \delta} \D^2 R| - \tfrac{5i}{8}(\D_{\alpha} \Db_{\dot \gamma} G_{\delta \dot \delta} + {\rm perms.})| - \tfrac{5}{4} \epsilon_{\alpha \delta} (\D^{\beta}_{\dot \gamma} G_{\beta \dot \delta} + \D^{\beta}_{\dot \delta}G_{\beta \dot \gamma})|\nonumber \ ,
\eea
where in the last line we used the fact that the anti-symmetric part in $\alpha$ and $\delta$ of the second term is zero. 
We therefore arrive at the following answer 
\begin{align}
(6) \ = \  \tfrac{i}{8}\, \epsilon^{\dalpha \dbeta}\,  \bar \sigma^{\delta \dot \delta}_a  \D_\alpha R_{\dalpha \delta \dot \delta \dgamma \dbeta}| \, \bar  \sigma^{\alpha \dot \gamma}_{b}\, \D^b \phid &\ =\ \Big(- \tfrac{9}{32} \D^2 R|\,  \eta_{ab} - \tfrac{5i}{32} \, \bar \sigma^{\delta \dot \delta}_a \bar \sigma^{\alpha \dgamma}_b \epsilon_{\alpha \delta} (\D^{\beta}_{\dot \gamma} G_{\beta \dot \delta} + \D^{\beta}_{\dot \delta}G_{\beta \dot \gamma})| \nonumber \\
 &\hspace{1cm} +\, \tfrac{5}{64} \bar \sigma_a^{\delta \dot \delta} \bar \sigma^{\alpha \dot \gamma}_{ b} (\D_{\alpha}\Db_{\dot \gamma} G_{\delta \dot \delta} + {\rm perms.})| \, \Big)\, \D^b \phid\ .  \label{equ:six}
  \end{align}
 We have written the result in terms of $G_a$, as this will combine with the next term to yield a simpler result. 

\vskip 4pt
- The seventh term has the following bosonic terms
\beq
(7) \equiv   \tfrac{1}{16} \epsilon^{\dalpha \dbeta} \D^2 \Db_{\dalpha} (T_{\dbeta a \dot \gamma}  \Db^{\dot \gamma} \Phid)| \ =\  \tfrac{1}{8}  \D^\alpha \Db^{\dbeta} T_{\dbeta a \dgamma}| \D_{\alpha} \Db^{\dgamma} \Phid| 
 - \tfrac{1}{32} \D^2 T^{\dbeta}_{\dbeta a}| \Db^2 \Phid| -  \tfrac{1}{32}  T^{\dbeta}_{\dbeta a}| \D^2 \Db^2 \Phid|  \ .
\eeq
Using $T^{\dbeta}_{\dbeta a} = 2 i G_a$ and $\tfrac{1}{16} \D^2 \Db^2 \Phid| = \hat \Box \phid + \tfrac{2}{3} \Md \Fd$, 
we find 
\begin{align}
(7a) &\ \equiv \ 
  - \tfrac{1}{32} \D^2 T^{\dbeta}_{\dbeta a}| \Db^2 \Phid| -  \tfrac{1}{32}  T^{\dbeta}_{\dbeta a}| \D^2 \Db^2 \Phid| \nonumber\\
  &\ =\  e^\mu_a \left[ \tfrac{i}{3} b_\mu \left(  \hat \Box \phid  + \tfrac{2}{3}\Md \Fd\right) + \tfrac{1}{6} \Fd \nabla_\mu  \Md + \tfrac{i}{6} b_\mu \Md \Fd \right]\ . \label{equ:7a}
\end{align}
To complete the calculation we need $ \D^\alpha \Db^{\dbeta} T_{\dbeta a \dgamma}|$.
First, we note that 
\bea
 \Db^{\dbeta} T_{\dbeta a \dgamma}& =&  \tfrac{i}{8} \bar \sigma^{\delta \dot \delta}_a\,\Db^{\dbeta} (-\epsilon_{ \dot \delta \dgamma} G_{\dbeta \delta} + 3 \epsilon_{\dbeta \dgamma} G_{\delta \dot \delta} + 3 \epsilon_{\dbeta \dot \delta} G_{\dot \gamma \delta} )  \nonumber \\
 &=& \tfrac{i}{8} \bar \sigma^{\delta \dot \delta}_a  (-\epsilon_{ \dot \delta \dgamma} \Db^{\dbeta} G_{\dbeta \delta} - 3 \Db_{\dgamma} G_{\delta \dot \delta} - 3\Db_{ \dot \delta} G_{\dot \gamma \delta} ) \nonumber \\
 &=&  \tfrac{i}{8} \bar \sigma^{\delta}_{ \dot \gamma a} \D_{\delta} R - \tfrac{3i}{8} (\Db_{\dgamma} G_{\delta \dot \delta} + \Db_{ \dot \delta} G_{\dot \gamma \delta} ) \bar \sigma^{\delta \dot \delta}_a\ .
\eea
This implies
\beq
\D^\alpha \Db^{\dbeta} T_{\dbeta a \dgamma} = \tfrac{i}{16}\delta^{\alpha}_{ \delta} \epsilon_{\dgamma \dot \delta} \bar \sigma^{\delta \dot \delta}_a \D^2  R - \tfrac{3}{8} \delta^{\alpha}_ {\delta}(\D^{\beta}_{\dgamma} G_{\beta \dot \delta} + \D^{\beta}_{\dot \delta} G_{\beta \dot \gamma}) \bar \sigma^{\delta \dot \delta}_a - \tfrac{3i}{16} \epsilon^{\alpha \beta} (\D_{\beta} \Db_{\dot \gamma} G_{\delta \dot \delta} + {\rm perms.}) \bar \sigma^{\delta \dot \delta}_a \ ,
\eeq
and hence
\bea
 (7b)  \equiv  \left(\tfrac{(-2i )}{8}\sigma^{\dgamma }_{\alpha b } \D^\alpha \bar \D^{\dbeta} T_{\dbeta a \dgamma}\right) 
 \D^b \Phid| &=& \left(\tfrac{1}{32}\D^2 R| \, \eta_{ab}   - \tfrac{3i}{32}  \bar \sigma^{\delta \dot \delta}_a \bar \sigma^{\alpha \dgamma }_b\epsilon_{\alpha \delta}(\D^{\beta}_{\dgamma} G_{\beta \dot \delta} + \D^{\beta}_{\dot \delta} G_{\beta \dot \gamma}) \right. \ \ \ \ \ \ \nonumber \\
 &&\ \ \left. +\, \tfrac{3}{64} \bar \sigma^{\delta \dot \delta}_a \bar \sigma^{\alpha \dgamma }_b  (\D_{\alpha} \Db_{\dot \gamma} G_{\delta \dot \delta} + {\rm perms.}) \right) 
 \D^b \phid \ . \label{equ:B35}
\eea
At this point, it makes sense to combine (\ref{equ:B35}) with (\ref{equ:six}),
\begin{align}
-(6) - (7b) 
&\ =\ \Big( \tfrac{8}{32} \D^2 R|\,  \eta_{ab}  + \tfrac{i}{4}  \bar \sigma^{\delta \dot \delta}_a \bar \sigma^{\alpha \dgamma}_b \epsilon_{\alpha \delta} (\D^{\beta}_{\dot \gamma} G_{\beta \dot \delta} + \D^{\beta}_{\dot \delta}G_{\beta \dot \gamma})| \nonumber \\
&\hspace{1cm} -  \tfrac{1}{8} \bar \sigma_a^{\delta \dot \delta} \bar \sigma^{\alpha \dot \gamma}_{ b} (\D_{\alpha}\Db_{\dot \gamma} G_{\delta \dot \delta} + {\rm perms.})| \,\Big) \, \D^b \phid  \ .
\end{align}
Using the contraction with $\sigma$ matrices
\bea
\bar \sigma^{\delta \dot \delta}_a \bar \sigma^{\alpha \dgamma}_b \epsilon_{\alpha \delta} (\D^{\beta}_{\dot \gamma} G_{\beta \dot \delta} + \D^{\beta}_{\dot \delta}G_{\beta \dot \gamma}) 
&=& 4 (\D_b G_a - \D_a G_b)+ 4 i \epsilon^{cd}{}_{ab} \D_c G_d \ , \label{eqn:sigmaidentity}\ 
\eea
and consulting eq.~(\ref{equ:A18}), we find
\begin{align}
-(6) - (7b) &\ =\ 
\Big( \left(\tfrac{1}{4} \D^2 R| + \tfrac{1}{2}(\tfrac{1}{2} {\cal R} + \tfrac{i}{3} \D_c b^c + \tfrac{1}{9} b_c b^c) \right) \eta_{ab}  \ \nonumber \\
 &\hspace{1cm} - \left( {\cal R}_{ab} + \tfrac{ 2i }{3} {\D}_b b_a + \tfrac{2}{9} b_a b_b\right)+ \tfrac{1}{3}\epsilon^{cd}{}_{ab} \D_c b_d \Big) \, \D^b \phid \, .
\end{align}
Combining this with (\ref{equ:7a}), we obtain
\bea
({\rm II}) \ \equiv \ -(6) - (7) &=& e^\mu_a \Big[  \left( \tfrac{1}{4} \D^2 R| + \tfrac{1}{2} (\tfrac{1}{2}{\cal R} + \tfrac{i}{3}{\nabla} \hskip -3pt \cdot \hskip -2pt b + \tfrac{1}{9} b^2) \right) \nabla_\mu \phid \nonumber \\
&& \hspace{0.7cm}  -\, \left( {\cal R}_{\mu \nu} + \tfrac{2 i}{3}{\nabla}_\nu b_\mu + \tfrac{2}{9} b_\mu b_\nu \right  )\nabla^\nu \phid +\tfrac{1}{3}\epsilon^{\rho \sigma}{}_{\mu \nu} \nabla_\rho b_\sigma \nabla^\nu \phid \nonumber \\
&& \hspace{0.7cm} - \ \tfrac{i}{3} b_\mu \bigl(  \hat \Box \phid  + \tfrac{2}{3}\Md \Fd\bigr) - \tfrac{1}{6} \Fd  \nabla_\mu  \Md - \tfrac{i}{6} b_\mu \Md \Fd \Big]\ .
\eea

\vskip 6pt
$\blacksquare$ FINALLY, we get, 
\begin{align}
\tfrac{1}{16}\D^2 \bar \D^2 \D_a \Phid| &\, =\, ({\rm I}) + ({\rm II}) \, =\,
\fbox{$\displaystyle e^{\mu}_a \left[\nabla_\mu \nabla^2 \phid + \tfrac{2i}{3} b^\nu \nabla_\mu \nabla_\nu \phid + \tfrac{1}{2} F_\mu{}^\nu \nabla_\nu \phid + F_\mu \right] $}\ ,
\end{align}
where
\bea
F_{\mu} &\equiv&  \tfrac{5}{6} \Md  \nabla_\mu \Fd + \tfrac{1}{3} \Fd \nabla_\mu \Md - \tfrac{i}{3} b_\mu \Md \Fd\ , \\
F_\mu{}^\nu &\equiv& \left(\tfrac{1}{6} {\cal R} + \tfrac{1}{6} |M|^2 + \tfrac{1}{3} b^2 - \tfrac{i}{3} \nabla \hskip -3pt \cdot \hskip -2pt b \right) \delta^\nu_\mu  \nonumber \\
&& +  \left( -2{\cal R}_{\mu}{}^{\nu} - \tfrac{4 i}{3} \nabla^\nu b_\mu +\tfrac{4i}{3} \nabla_\mu b^\nu - \tfrac{4}{9}  b_\mu b^\nu  + \tfrac{2}{3} \epsilon^{\rho \sigma}{}_\mu{}^\nu \nabla_\rho b_\sigma \right) \, . 
\eea

\newpage
 \begingroup\raggedright\endgroup

\end{document}